\ifdefined\CameraReadyMain
  \documentclass[journal]{vgtc}                        % camera-ready main paper
\else
  \documentclass[journal]{vgtc}                        % integrated paper or standalone appendix
\fi
\onlineid{1580}

\vgtccategory{Research}

\ifdefined\AppendixOnly
  \title{Supplemental Appendix for DataMagic: Authoring Data Videos through Declarative Multi-Agent Orchestration}
\else
  \title{DataMagic: Authoring Data Videos through Declarative Multi-Agent Orchestration}
\fi
\author{%
  \authororcid{Yupeng Xie}{0009-0002-6489-7171},
  \authororcid{Zhenyang Wang}{0009-0001-3513-6763},
  \authororcid{Liangwei Wang}{0000-0003-3481-3993},
  \authororcid{Jiayi Zhu}{0009-0005-5537-7408},
  \authororcid{Zhouan Shen}{0009-0006-8209-9842},
  and \authororcid{Yuyu Luo}{0000-0001-9530-3327}%
}

\authorfooter{
  \item Yupeng Xie, Zhenyang Wang, Liangwei Wang, Jiayi Zhu, Zhouan Shen,
    and Yuyu Luo are with The Hong Kong University of Science and Technology
    (Guangzhou), Guangzhou, China. E-mail: yxie740@connect.hkust-gz.edu.cn,
    wwwangzhenyang@gmail.com, lwang344@connect.hkust-gz.edu.cn,
    jzhu351@connect.hkust-gz.edu.cn, zshen575@connect.hkust-gz.edu.cn,
    and yuyuluo@hkust-gz.edu.cn. Yuyu Luo is the corresponding author.
}

\ifdefined\AppendixOnly
\abstract{}
\else
\abstract{%
  Data videos communicate data insights through dynamic charts, voice narration, and synchronized animations, and have become a widely adopted form of data storytelling. However, their production requires multidisciplinary expertise spanning data analysis, narrative design, and video editing. Static visualization tools lack narrative and animation capabilities; authoring tools rely on pre-prepared charts rather than raw data; and pixel-level generation models, while capable of end-to-end synthesis, cannot guarantee data accuracy or provenance. End-to-end automatic generation faces two core challenges: how to uniformly represent charts, narration, and animations together with their temporal relationships, and how to efficiently search a vast design space for narrative-coherent compositions.
  We present {\sc DataMagic}, a system that authors data videos from raw tabular data through declarative multi-agent orchestration, built on two core designs. First, the declarative specification {\sc DVSpec} unifies charts, narration, and animations with data-bound references and declarative synchronization, ensuring data provenance and automatic audio-visual alignment. Second, a ``Generate-then-Orchestrate'' multi-agent strategy generates candidate scenes in parallel and then optimizes narrative coherence through global orchestration. {\sc DVSpec} further serves as a shared state supporting three complementary interaction modes, bridging full automation with fine-grained human control.
  Evaluations on 109 real-world samples show that even the most advanced LLM (e.g., GPT-5) achieves only 2.13/5 with execution success rates between 48.62\% and 86.24\%; {\sc DataMagic} improves quality to 3.89 (+83\%) with success rates above 95\%, with the most significant gains in animation and narrative dimensions. A user study further demonstrates that, compared to a conversational LLM workflow, {\sc DataMagic} significantly improves creation efficiency (79.7\% reduction in task time) and reduces perceived cognitive load. Source code is available at \url{https://github.com/HKUSTDial/DataMagic}.%
}

\fi

\newif\ifshowack
\showacktrue

\graphicspath{{figures/}{reference/autodv/figures/}{figs/}{./}}

\usepackage{mathptmx}
\usepackage{amsmath}
\usepackage{amssymb}
\usepackage{mathtools}
\usepackage{amsthm}

\usepackage{microtype}
\usepackage{booktabs}
\usepackage{multirow}
\usepackage{makecell}
\usepackage{pifont}

\usepackage{graphicx}
\usepackage{subcaption}
\usepackage{placeins}

\usepackage{algorithm}
\usepackage{algorithmic}

\usepackage{xcolor}
\usepackage{xspace}
\usepackage{listings}
\lstdefinestyle{jsonstyle}{
  breaklines=true,
  breakatwhitespace=true,
  columns=fullflexible,
  basicstyle=\ttfamily\small,
  showstringspaces=false,
  numbers=none,
  frame=none
}
\lstdefinestyle{typescriptstyle}{
  breaklines=true,
  breakatwhitespace=true,
  columns=fullflexible,
  basicstyle=\ttfamily\small,
  showstringspaces=false,
  numbers=none,
  frame=none,
  keywordstyle=\color{blue!70},
  commentstyle=\color{green!50!black}
}
\usepackage{tcolorbox}
\tcbuselibrary{listings,breakable}
\definecolor{framegray}{HTML}{595959}
\definecolor{backgray}{HTML}{F2F2F2}
\newtcolorbox{examplebox}[1]{
  width=\linewidth,
  breakable,
  break at=-\baselineskip/0pt,
  colback=backgray,
  colframe=framegray,
  boxrule=0.5pt,
  leftrule=2pt,
  toprule=1pt,
  bottomrule=2pt,
  rightrule=2pt,
  arc=0pt,
  title={\textbf{#1}},
  label={box:#1},
  fonttitle=\bfseries,
  before skip=0.5\baselineskip,
  after skip=0.5\baselineskip,
  enforce breakable
}

\newcommand{\system}{{\sc DataMagic}\xspace}
\newcommand{\grammar}{{\sc DVSpec}\xspace}
\newcommand{\suppappendixref}[2]{%
  \ifdefined\CameraReadyMain
    Appendix~#1 of the supplemental material%
  \else
    Appendix~\ref{#2}%
  \fi
}
\newcommand{\suppappendixsubref}[2]{%
  \ifdefined\CameraReadyMain
    #1%
  \else
    Appendix~\ref{#2}%
  \fi
}
\newcommand{\evaluationappendixstatement}{%
  \ifdefined\CameraReadyMain
    The supplemental material provides scoring prompts in Appendix~D and dimension-level consistency results and expert-validation interface and procedure details in Appendix~E.%
  \else
    Detailed scoring prompts are provided in Appendix~\ref{appendix:detailed_evaluation_questions}, and dimension-level consistency results together with details of the expert-validation interface and procedure are provided in Appendix~\ref{appendix:expert_validation}.%
  \fi
}

\definecolor{mygreen}{RGB}{0,128,0}

\newcommand{\rev}[1]{#1}
\theoremstyle{plain}

\theoremstyle{definition}

\theoremstyle{remark}

\begin{document}

\ifdefined\AppendixOnly

\renewcommand{\manuscriptnotetxt}{}
\maketitle

\appendix
\crefalias{section}{appendix}
\setcounter{figure}{0}
\setcounter{table}{0}

\section{Case Studies}
\label{appendix:case_studies}

\begin{figure*}[!t]
    \centering
    \includegraphics[width=1\linewidth]{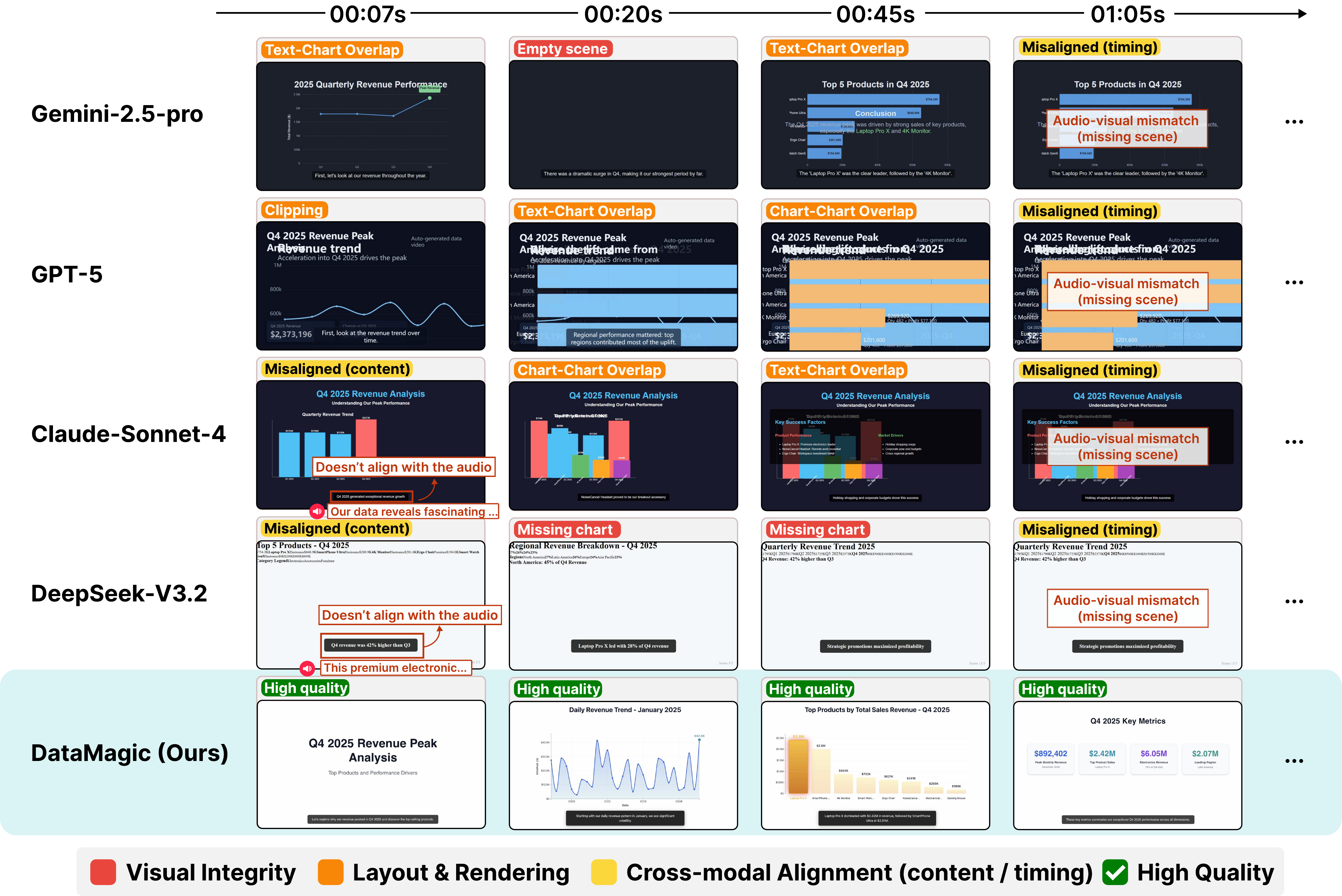}
    % \vspace{-0.5em}
    \caption{\rev{Comparative Analysis of Video Generation Quality Across Different Methods}}
    \label{fig:case_study}
    % \vspace{-1em}
\end{figure*}

\rev{This appendix presents representative qualitative cases for \system and direct-generation baselines, covering cross-method comparisons, high-quality generations, and failure-mode analyses across complex data storytelling scenarios such as agriculture, environmental science, business analytics, energy, and gender-based compensation.}

% --- A.1 Cross-Method Comparison ---
\subsection{\rev{Cross-Method Comparison}}
\label{subsec:appendix_comparison}

\noindent\rev{\textbf{Comparative Case.} As shown in Figure~\ref{fig:case_study}, using the same input conditions (Q4 2025 revenue analysis query and transaction dataset), direct generation methods frequently exhibit all three error categories: empty scenes or missing charts (visual rendering failures), visual element overlap (design defects), and narration--visual content or timing mismatches (audio-visual errors). In contrast, \system generates well-structured videos with precise audio-visual alignment, achieving an average score of 4.2 (vs.\ 1.5 for direct generation) with consistent improvements across all evaluation dimensions.}

% --- A.2 High-Quality Generations ---
\subsection{\rev{High-Quality Generations}}
\label{subsec:case_study}

\begin{figure*}[!t]
    \centering
    \includegraphics[width=1\linewidth]{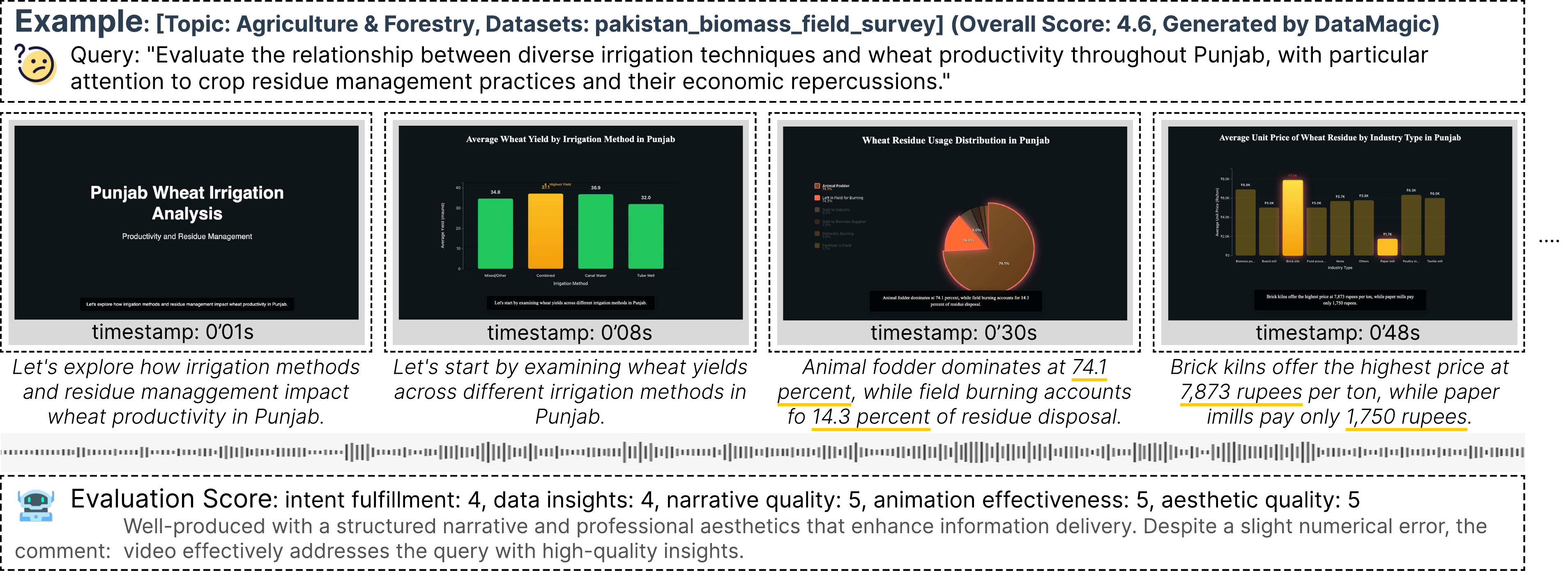}
    % \vspace{-0.5em}
    \caption{\rev{A representative high-quality case generated by \system (Agriculture, overall score: 4.6). Four key frames are shown with their corresponding narration (italicized). Underlined spans in the narration are the index triggers in \grammar that fire the co-occurring visual highlights, demonstrating precise audio-visual synchronization without manual timeline adjustment.}}
    \label{fig:high_case_1}
    % \vspace{-1.5em}
\end{figure*}

\noindent\rev{\textbf{Representative Agriculture Case.} Figure~\ref{fig:high_case_1} presents a representative high-quality generation on a real-world agriculture dataset (overall score: 4.6). Given the query \textit{``Evaluate the relationship between diverse irrigation techniques and wheat productivity throughout Punjab, with particular attention to crop residue management practices and their economic repercussions,''} \system constructs a coherent four-scene narrative: an overview introduction, followed by wheat-yield comparisons across irrigation methods (0'08s), residue-disposal distributions (0'30s), and industry-level price disparities (0'48s). Within the residue-disposal scene, when the narrator states \textit{``animal fodder dominates at 74.1\%,''} the corresponding pie-chart segment is highlighted simultaneously, demonstrating narration-synchronized cross-modal alignment without manual timeline adjustment. Both narrative quality and animation effectiveness score 5/5.}

% --- Additional High Quality Examples ---
\label{subsec:high_quality_cases}

\begin{figure*}[t!]
    \centering
    \includegraphics[width=1\linewidth]{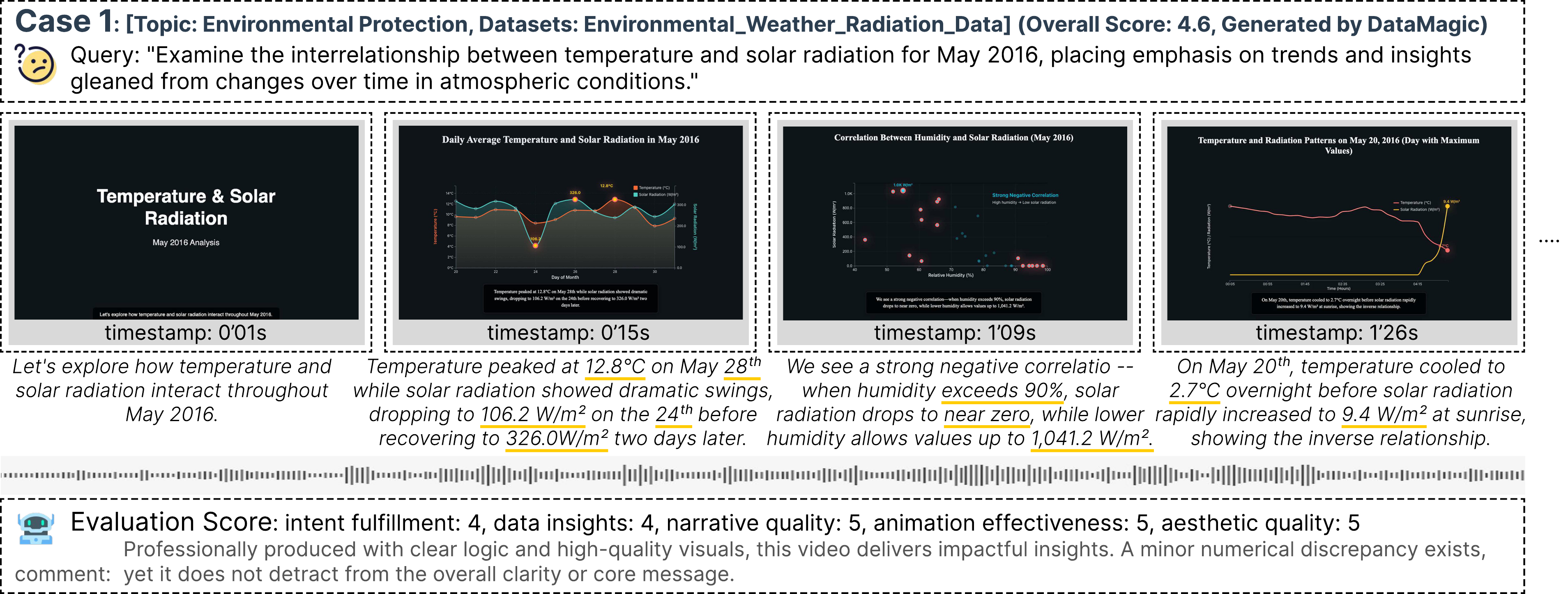}
    \caption{High-quality Case 1 (Environmental Science): High-fidelity tracking of meteorological fluctuations and solar radiation swings.}
    \label{fig:high_case_2}
\end{figure*}

\begin{figure*}[t!]
    \centering
    \includegraphics[width=1\linewidth]{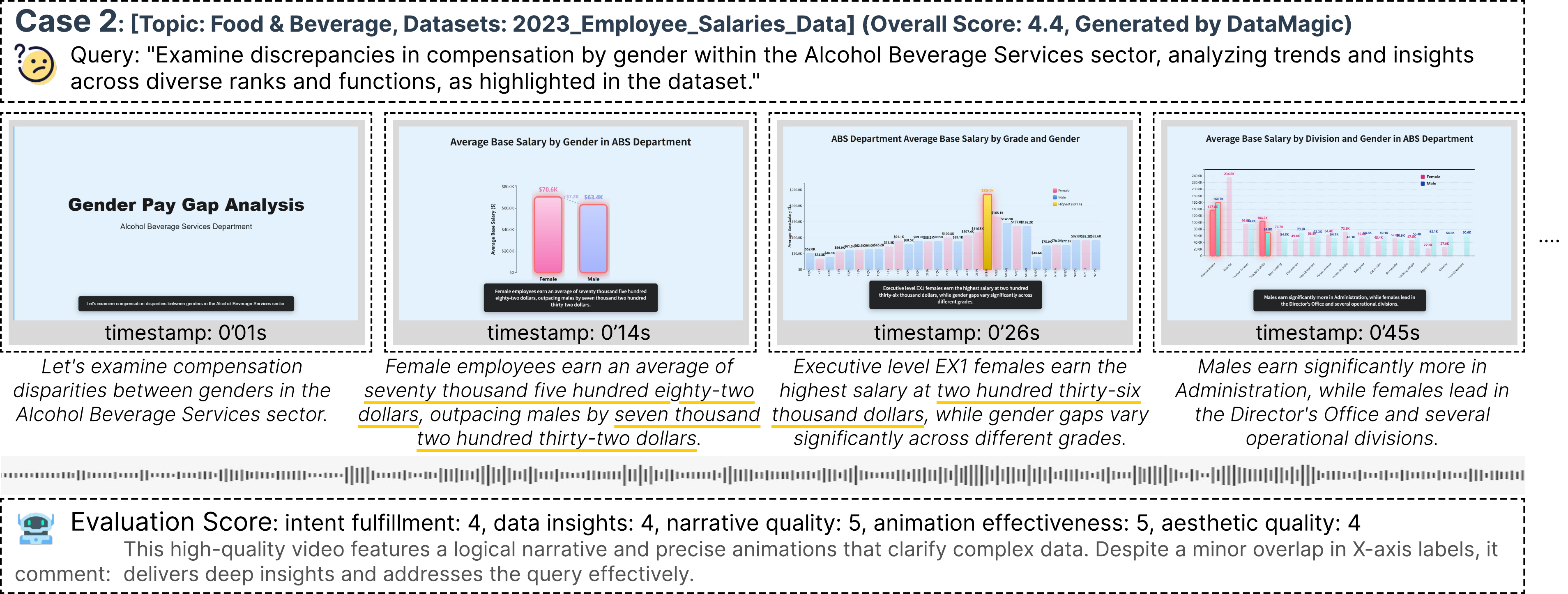}
    \caption{High-quality Case 2 (Gender-based Compensation): Visualization of complex gender-based compensation disparities across divisions.}
    \label{fig:high_case_3}
\end{figure*}

\noindent\rev{\textbf{Additional Cross-Domain Examples.} Figures~\ref{fig:high_case_2} and~\ref{fig:high_case_3} provide two further high-quality cases from environmental science and gender-based compensation, illustrating \system's ability to preserve numerical fidelity, convey fine-grained comparisons, and maintain coherent narration--visual coordination across domains.}

\noindent\textbf{Numerical Accuracy in Scientific Data (Environmental Science).} 
The sensitivity to data fluctuations is evident in the environmental science case (Figure~\ref{fig:high_case_2}). When analyzing meteorological data from May 2016, the system correctly identifies the temperature peak of 12.8°C on May 28th and synchronizes the visualization of solar radiation as it recovers from 106.2 W/m² to 326.0 W/m² within two days. Through the logical progression in Charts 3 and 4, the system reveals the inverse relationship between variables, such as radiation dropping to near zero when humidity exceeds 90\%, and radiation increasing to 9.4 W/m² as temperatures cool to 2.7°C overnight.

\noindent\textbf{Representation of Fine-Grained Disparities (Gender-based Compensation).} 
Furthermore, the gender-based compensation case (Figure~\ref{fig:high_case_3}) validates the ability of the system to represent fine-grained compensation disparities. In Chart 2, the visual output effectively supports the narrative's finding that female employees earn an average of \$70,582, surpassing males by \$7,232. To provide deeper insights, Chart 3 identifies the highest salary point for EX1 level females at \$236,000, while Chart 4 utilizes departmental comparisons to clarify the distribution of gender pay gaps across administration and operational divisions.

% --- A.3 Low Quality Examples ---
\subsection{Analysis of Low-Quality Generations}
\label{subsec:low_quality_cases}

Despite the robustness of \system, certain limitations persist when dealing with extreme data distributions or complex multi-modal instructions.

\begin{figure*}[t!]
    \centering
    \includegraphics[width=1\linewidth]{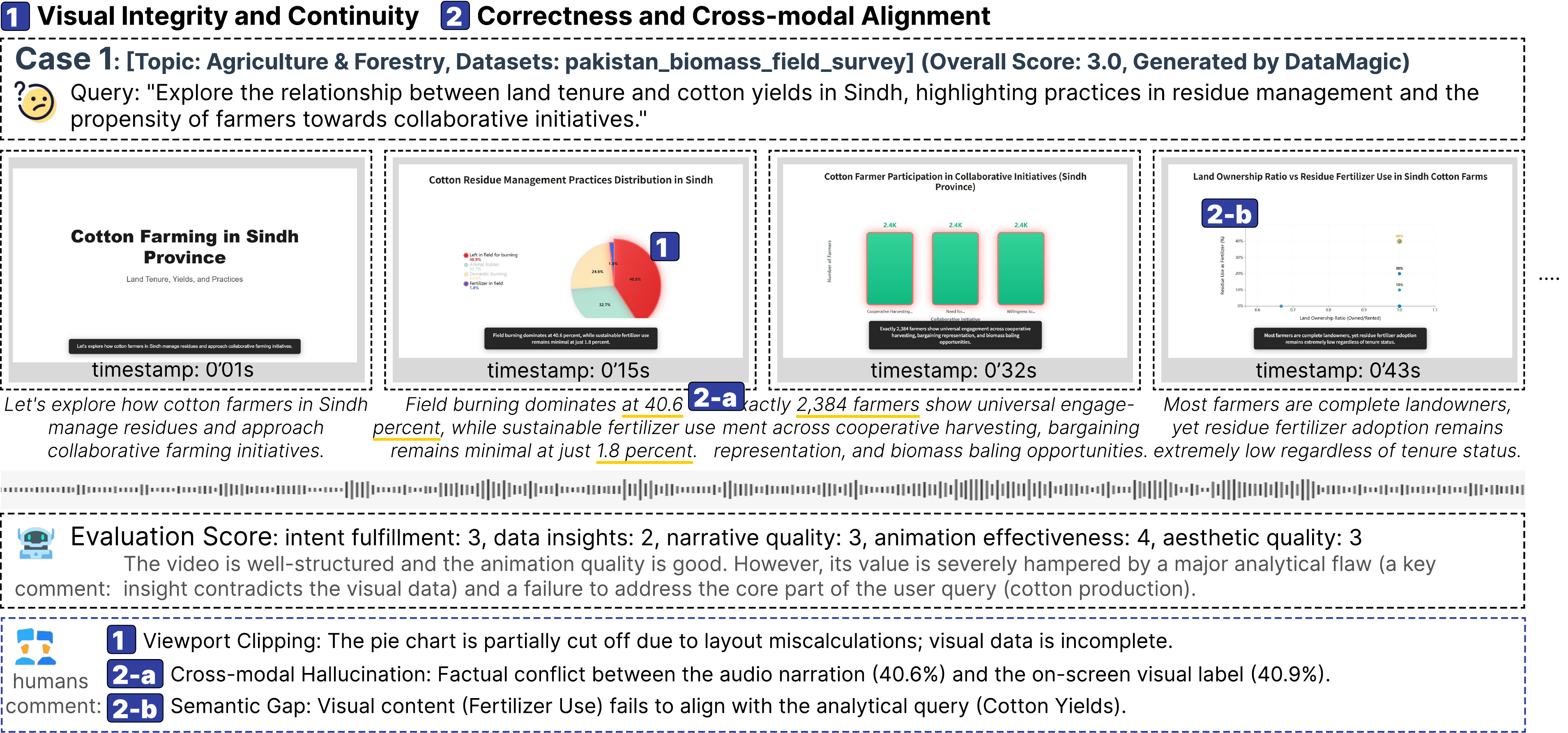}
    \caption{Low-quality Case 1 (Agriculture): Analysis of layout truncation and numerical discrepancies in \system under extreme data distributions.}
    \label{fig:low_quality_case_1}
\end{figure*}

\noindent\textbf{Analysis of \system Limitations (Agriculture).}
In the agriculture case shown in Figure~\ref{fig:low_quality_case_1}, extreme disparities in input values cause the visualization elements in Chart 2 to fail to adapt to the preset layout, resulting in Viewport Clipping [1] where the pie chart is partially truncated. This instability is also reflected in cross-modal consistency; for instance, a Cross-modal Hallucination [2-a] occurs where the audio narration mentions a value of 40.6\% while the visual label displays 40.9\%. Additionally, Chart 4 reveals a Semantic Gap [2-b] where the visual content focuses on land ownership and fertilizer use despite the query specifically requesting an analysis of cotton yields, indicating a failure to fully address the user's analytical intent.

\begin{figure*}[t!]
    \centering
    \includegraphics[width=1\linewidth]{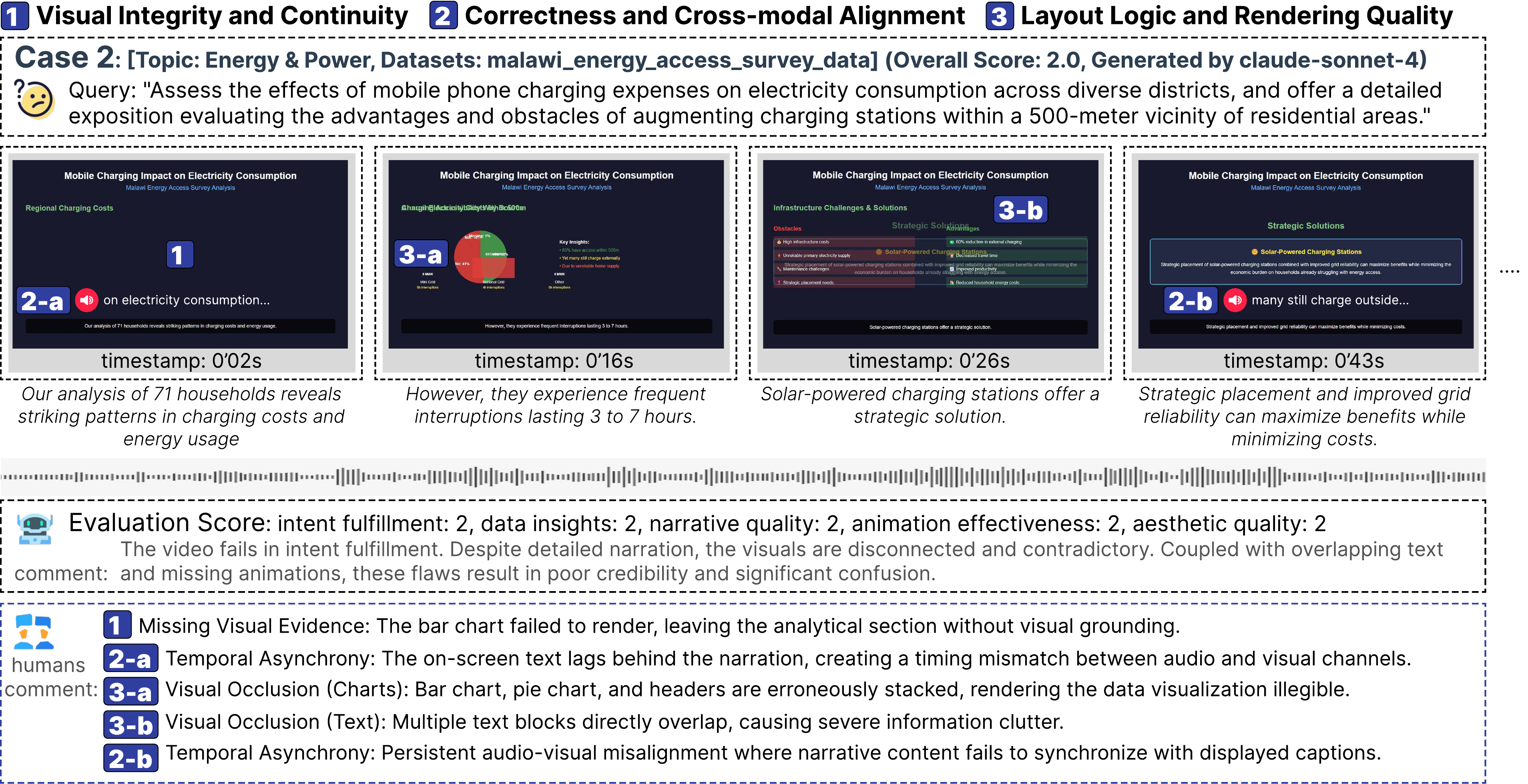}
    \caption{Low-quality Case 2 (Energy): Analysis of systemic failures in baseline models across layout logic and cross-modal alignment.}
    \label{fig:low_quality_case_2}
\end{figure*}

\noindent\textbf{Analysis of Baseline Limitations (Energy).}
By comparison, the limitations of the baseline model (based on Claude-Sonnet-4) are more pronounced in the energy analysis case (Figure~\ref{fig:low_quality_case_2}). The baseline fails in Chart 1 due to Missing Visual Evidence [1], where the bar chart is not rendered, accompanied by severe Temporal Asynchrony [2-a] where the text lags behind the audio. In Charts 2 and 3, a lack of layout logic leads to significant Visual Occlusion, characterized by the erroneous stacking of multiple charts [3-a] and overlapping text blocks for strategic solutions [3-b], rendering the information illegible. Finally, persistent Temporal Asynchrony [2-b] in Chart 4 results in a complete disconnect between the captions and the narration, hindering the overall coherence of the information delivery.

\section{Evaluation Dataset Statistics}
\label{app:dataset}

We build our evaluation set based on two representative benchmark datasets: T2R-bench~\cite{zhang2025t2r}  and DAComp-DA~\cite{lei2025dacomp}. The final evaluation set contains 60 datasets and 109 test samples, covering diverse data scales, query types, and application domains.

\subsection{Dataset Scale Distribution and Dimensions}

The dataset sources and scale statistics are shown in Table~\ref{tab:dataset_source_scale}. Overall, our evaluation set covers a broad range of dataset sizes, from fewer than 100 rows (Small) to more than 1K rows (Large), sufficient for evaluating \system across different data volumes. Table~\ref{tab:dataset_dimension_stats} further summarizes the physical dimensions of the 60 datasets: the largest dataset contains 150{,}000 rows, and some tables contain up to 555 columns, with numeric columns dominating (up to 552) and categorical columns contributing additional semantic signals (up to 57).

\begin{table}[t!]
  \caption{Dataset sources and scale distribution.}
  \label{tab:dataset_source_scale}
  \centering
  \begin{small}
    
    \setlength{\tabcolsep}{3pt}   
      \begin{tabular}{lccccc}
        \toprule
        \textbf{Source} & \textbf{Datasets} & \textbf{Samples} &  
        \textbf{\shortstack[c]{Small\\(<100 rows)}} &                    
        \textbf{\shortstack[c]{Med.\\(100-1K)}} &                   
        \textbf{\shortstack[c]{Large\\(>1K rows)}} \\                    
        \midrule
        T2R-bench  & 36 & 85  & 4 & 14 & 18 \\
        DAComp-DA  & 24 & 24  & 2 & 7  & 15 \\
        \midrule
        Total      & 60 & 109 & 6 & 21 & 33 \\
        \bottomrule
      \end{tabular}
    
  \end{small}
\end{table}

\begin{table}[t!]
\centering
\caption{Dataset dimension statistics.}
\label{tab:dataset_dimension_stats}
\renewcommand{\arraystretch}{1.15}
\setlength{\tabcolsep}{4pt}
\small
\begin{tabular}{lcccc}
\toprule
\textbf{Dimension} & \textbf{Min} & \textbf{Median} & \textbf{Max} & \textbf{Mean} \\
\midrule
Rows & 23 & 1{,}436 & 150{,}000 & 17{,}065.58 \\
Columns & 3 & 16 & 555 & 32.42 \\
\quad Numeric columns & 1 & 8 & 552 & 24.25 \\
\quad Categorical columns & 0 & 7 & 57 & 8.17 \\
\bottomrule
\end{tabular}
\end{table}

\subsection{Query Type Analysis}

The query type distribution of the 109 test samples is summarized in Table~\ref{tab:query_type_stats}, showing a broad coverage of data analysis tasks from multiple perspectives. The comprehensive analysis represents more than one quarter of the samples, which increases the complexity and difficulty of the task. Meanwhile, correlation analysis, trend analysis, comparative analysis, and distribution analysis provide a balanced task structure, ensuring that the evaluation set can thoroughly assess \system's logical reasoning and narrative quality in diverse data video composition scenarios.

\begin{table}[t!]
\centering
\caption{Query type distribution of the sample dataset.}
\label{tab:query_type_stats}
\renewcommand{\arraystretch}{1.15}
\setlength{\tabcolsep}{3pt}
\small
\begin{tabular}{lccl}
\toprule
\textbf{Query Type} & \textbf{Count} & \textbf{Percentage} & \textbf{Typical Patterns} \\
\midrule
\makecell[l]{Trend analysis}
& \makecell[c]{12}
& \makecell[c]{11.0\%}
& \makecell[l]{Time-series changes;\\Periodic patterns} \\

\makecell[l]{Comparative analysis}
& \makecell[c]{15}
& \makecell[c]{13.8\%}
& \makecell[l]{Cross-category comparison;\\Ranking analysis} \\

\makecell[l]{Distribution analysis}
& \makecell[c]{18}
& \makecell[c]{16.5\%}
& \makecell[l]{Value distribution;\\Proportion} \\

\makecell[l]{Correlation analysis}
& \makecell[c]{33}
& \makecell[c]{30.2\%}
& \makecell[l]{Variable relationships;\\Correlation} \\

\makecell[l]{Comprehensive analysis}
& \makecell[c]{31}
& \makecell[c]{28.4\%}
& \makecell[l]{Multi-dimensional analysis} \\
\midrule
\textbf{Total} & \textbf{109} & \textbf{100.0\%} & -- \\
\bottomrule
\end{tabular}
\end{table}

\subsection{Domain and subdomain of evaluation data}

By combining real-world industrial data from T2R-bench with enterprise business data from DAComp-DA, we summarize the evaluation set into 6 domains and 22 sub-domains, as shown in Table~\ref{tab:domain_subdomain}. This setting provides broad domain coverage and matches real-world application needs.

\begin{table}[t!]
\centering
\caption{Domains and sub-domains.}
\label{tab:domain_subdomain}
\renewcommand{\arraystretch}{1.15}
\setlength{\tabcolsep}{5pt}
\begin{small}
\begin{tabular}{ll}
\toprule
\textbf{Domains} & \textbf{Sub-domains} \\
\midrule
\makecell[l]{Technology and \\ Engineering} &
\makecell[l]{Electronics and Automation Manufacturing;\\
Academic Research; \\
Energy Production and Power Systems;\\
Automotive Industry} \\
\midrule
\makecell[l]{Environmental\\Management} &
\makecell[l]{Environmental Protection;\\
Agriculture and Forestry;\\
Resource Management} \\

\midrule
\makecell[l]{Transportation\\Logistics} &
\makecell[l]{Communication and Digital Infrastructure;\\
Transportation Networks and Logistics Management} \\
\midrule
\makecell[l]{Social Policy\\Administration} &
\makecell[l]{Education Policy and Public Education;\\
Government Administration and Public Sector Services;\\
Labor and Employment Administration; \\
Healthcare Systems and Public Health;\\
Demographics and Social Development} \\
\midrule
\makecell[l]{Commercial Services \\ and Markets} &
\makecell[l]{Retail Trade and E-commerce Platforms;\\
Tourism and Hospitality Services;\\
Digital Entertainment and Gaming; \\
Food and Beverage Services;\\
Real Estate and Housing Market; \\
Business Management and Supply Chain} \\
\midrule
\makecell[l]{Financial Economics} &
\makecell[l]{Economic Development and International Trade;\\
Banking and Financial Services} \\
\bottomrule
\end{tabular}
\end{small}
\end{table}

\section{Agent Prompts}
\label{appendix:agent_prompts}

\system’s multi-agent framework relies on carefully crafted system prompts to enforce clear boundaries of responsibility, standardized input–output formats, and explicit collaboration protocols across agents. In this appendix, we present the prompt structure and core instructions for the key agents.

There are two stages in the pipeline. The first stage is to generate configuration files for the video. The second stage is to generate executable files based on the configuration files generated in the first stage.

\textbf{Note:} In the following prompts, metadata refers to the summary of the dataset structure, not the full data.

\subsection{Scene Planner Agent}
\label{appendix:scene_planner_prompt}

\begin{examplebox}{Prompt Template: Scene Planner Agent}
You are a data analysis planner specializing in planning visualization analyses.\\

\textbf{Task}: Based on the user's query and dataset summary, plan what data analyses are needed to answer the query.

\subsubsection*{\textbf{Input}}
\begin{itemize}
    \item \textbf{User Query}: \{query\}
    \item \textbf{Dataset Summary}: \{metadata\}
\end{itemize}

\subsubsection*{\textbf{Your Role}}
Plan the material list - what analyses/visualizations should be created. Think like a photographer planning what shots to take, NOT like a director planning the final edit sequence.

\subsubsection*{\textbf{Output Format (JSON)}}
\begin{lstlisting}[style=jsonstyle]
{
  "narrative_pattern": "freytag_default",
  "scenes": [
    {
      "id": "unique_id",
      "type": "chart",
      "query": "Specific query with column names",
      "analysis_type": "primary_intent",
      "intent": ["intent1", "intent2"],
      "priority": 0.0-1.0,
      "context": "Why this analysis matters",
      "required_fields": ["field1", "field2"]
    }
  ]
}
\end{lstlisting}

\subsubsection*{\textbf{Analysis Types}}
trend, composition, comparison, distribution, correlation, find\_extremum, rank, summary, magnitude.

\subsubsection*{\textbf{Narrative Pattern}}
Choose one pattern that best fits the \emph{shape} of the story, used later by the Narrative Director to sequence scenes: \texttt{freytag\_default} (general-purpose, single dramatic peak; the default), \texttt{hook\_then\_evidence} (news-style inverted pyramid: the most striking finding stated first), \texttt{comparison\_driven} (A-vs-B contrasts dominate), \texttt{time\_driven} (chronological progression), \texttt{drill\_down} (overview to segments to individual drivers).

\subsubsection*{\textbf{Critical Rules}}
\begin{itemize}
    \item Plan ONLY \texttt{``chart''} type scenes (stat\_cards generated automatically later)
    \item Each scene must have specific, executable query with explicit column names
    \item Use priority (0.0-1.0): 1.0 = main query answer, 0.8-0.9 = key supporting, 0.6-0.7 = context, 0.4-0.5 = details, 0.2-0.3 = optional
    \item Provide context explaining why each analysis matters
    \item Focus on WHAT to analyze, not HOW to visualize
    \item Select 2-4 distinct angles, ensure adjacent scenes have different intent lists
    \item Avoid redundant scenes or multiple scenes showing same data from nearly identical angles
    \item Output a top-level \texttt{narrative\_pattern} (one of the five values above) based on the shape of the story
    \item DO NOT include: opening/closing scenes, stat cards, strict scene order, narration text, chart type specifications
\end{itemize}

\subsubsection*{\textbf{Examples}}
\{example\}

\subsubsection*{\textbf{Final Instruction}}
Now create the analysis plan. Return ONLY the JSON, nothing else.

\end{examplebox}

\textbf{Example:} The following demonstrates a concrete example of the Scene Planner Agent's output format and decision-making process. This example illustrates how to structure the analysis plan based on a user query.

\subsubsection{Scene Planner Agent Example}
\label{scene-planner-example}
\begin{examplebox}
    {Query: ``Compare flight delays across carriers and destinations in 2015''}
    
    \begin{lstlisting}[style=jsonstyle]
  "scenes": [
    {
      "id": "analysis_carrier_comparison",
      "type": "chart",
      "query": "Group by carrier column, calculate average of depdelay column",
      "analysis_type": "comparison",
      "priority": 1.0,
      "context": "Primary analysis: carrier performance comparison"
    },
    {
      "id": "analysis_destination_delays",
      "type": "chart",
      "query": "Group by destcity column, calculate average of arrdelay column",
      "analysis_type": "distribution",
      "priority": 0.9,
      "context": "Primary analysis: destination city delays"
    }
  ]
    \end{lstlisting}
    \textbf{Note}: No stat\_cards needed - charts clearly show the comparisons requested.
\end{examplebox}

\subsection{Data Preparation Agent}

\begin{examplebox}{Prompt Template: Data Preparation Agent}
You are a data transformation planner.\\

\textbf{Task}: Analyze the sub-query and determine what data transformations are needed to prepare the data for visualization.

\subsubsection*{\textbf{Input}}
\begin{itemize}
    \item \textbf{Sub-Query}: \{sub\_query\}
    \item \textbf{Analysis Type}: \{analysis\_type\}
    \item \textbf{Required Fields}: \{required\_fields\}
    \item \textbf{Dataset Metadata}: \{metadata\}
\end{itemize}

\subsubsection*{\textbf{Transformation Types}}
\begin{itemize}
    \item \texttt{group\_by\_aggregate}: Comparisons, distributions, part-to-whole, overall statistics, or 2D heatmap (use TWO fields in \texttt{group\_by\_fields})
    \item \texttt{time\_series\_aggregate}: Trends over time
    \item \texttt{top\_n}: Finding extremes, rankings
    \item \texttt{correlation\_data}: Correlation analysis, scatter plots (ALWAYS use this for correlation, NOT \texttt{filter\_and\_select})
    \item \texttt{filter\_and\_select}: Simple filtering and field selection
    \item \texttt{sample\_representative}: General analysis when no specific aggregation needed
\end{itemize}

\subsubsection*{\textbf{Output Format (JSON)}}
\begin{lstlisting}[style=jsonstyle, breaklines=true, breakatwhitespace=false]
{
  "transformation_type":"group_by_aggregate|time_series_aggregate|filter_and_select|top_n|correlation_data|sample_representative",
  "group_by_fields": ["field1"],
  "aggregate_fields": {"field2": "avg", "field3": "sum"},
  "time_field": "date",
  "time_grouping": "month",
  "x_field": "field1",
  "y_field": "field2",
  "derived_fields": {"new_field": "field1 / field2"},
  "filter": {"field": "value", "field2": "not null"},
  "select_fields": ["field1", "field2"],
  "sort_by": "avg_field2",
  "sort_order": "desc",
  "limit": 10,
  "sample_size": 20,
  "description": "Brief description of what this transformation does"
}
\end{lstlisting}

\subsubsection*{\textbf{Critical Rules}}
\begin{itemize}
    \item Use ONLY existing fields from the dataset metadata
    \item Determine ALL necessary filters from sub-query (scope, quality, logical)
    \item For overall statistics: use \texttt{group\_by\_aggregate} with empty \texttt{group\_by\_fields}
    \item Keep filters minimal - only add what's explicitly required
    \item Use actual values in filters, NO placeholders
    \item If query mentions specific location, ALWAYS include filter field
    \item Always include a clear description
    \item Return a SINGLE JSON OBJECT, NOT an array
\end{itemize}

\subsubsection*{\textbf{Examples}}
\{example\}

\subsubsection*{\textbf{Final Instruction}}
Now plan the transformation. Return ONLY the JSON object, nothing else.

\end{examplebox}

\textbf{Example:} The following demonstrates a concrete example of the Data Preparation Agent's transformation planning process. This example shows how to analyze a sub-query and determine the appropriate data transformation with filters and aggregation.

\subsubsection{Data Preparation Agent Example}
\label{data-preparation-example}

\begin{examplebox}
    {Comparison with Scope Filter}
    \textbf{Sub-query}:``Compare wheat yield across different irrigation methods in Punjab province''\\
    \textbf{Analysis Type}: comparison\\
    \textbf{Required Fields}: ["Province", "Crop", "Method of irrigation", "Yield, maund"]\\
    \textbf{Output}:
    \begin{lstlisting}[style=jsonstyle]
{
  "transformation_type": "group_by_aggregate",
  "group_by_fields": ["Method of irrigation"],
  "aggregate_fields": {
    "Yield, maund": "avg"
  },
  "filter": {
    "Province": "Punjab",
    "Crop": "Wheat",
    "Yield, maund": "not null",
    "Method of irrigation": "not null"
  },
  "sort_by": "avg_Yield, maund",
  "sort_order": "desc",
  "description": "Filter to Punjab wheat data, group by irrigation method, calculate average yield excluding null values"
}
    \end{lstlisting}
\end{examplebox}

\subsection{Visual Designer Agent}
\begin{examplebox}
{Prompt Template: Visual Designer Agent}
You are a professional data visualization designer specializing in chart configuration.\\

\textbf{Task}: Generate a chart visualization configuration based on the query and data. DO NOT write narration - only provide a brief insight summary.

\subsubsection*{\textbf{Input}}
\begin{itemize}
    \item \textbf{Sub-Query}: \{query\}
    \item \textbf{Analysis Type}: \{analysis\_type\}
    \item \textbf{Planned Scene Type}: \{planned\_type\} (MUST follow this type)
    \item \textbf{Processed Data}: \{data\}
\end{itemize}

\subsubsection*{\textbf{Chart Selection \& Data Requirements}}
\begin{itemize}
    \item Select appropriate chart type based on analysis type and data characteristics:
    \begin{itemize}
        \item \texttt{comparison/magnitude} $\rightarrow$ bar\_chart, grouped\_bar\_chart ($\geq$2 data points)
        \item \texttt{trend/change\_over\_time} $\rightarrow$ line\_chart, area\_chart ($\geq$3 time points)
        \item \texttt{part\_to\_whole/proportion} $\rightarrow$ pie\_chart, donut\_chart ($\geq$2 categories)
        \item \texttt{hierarchical/composition} $\rightarrow$ treemap, sunburst ($\geq$3 hierarchical nodes)
        \item \texttt{correlation/distribution} $\rightarrow$ scatter\_chart, bubble\_chart ($\geq$3 data points)
        \item \texttt{2D distribution/density} $\rightarrow$ heatmap, contour\_plot ($\geq$4 data points)
        \item \texttt{ranking/ordered\_comparison} $\rightarrow$ horizontal\_bar\_chart, lollipop\_chart
    \end{itemize}
    \item \textbf{Note}: The chart types above represent commonly used configurations. The declarative specification framework supports additional visualization types through the extensible rendering engine.
    \item If insufficient data for charts, create stat\_cards instead
    \item For \texttt{``stat\_cards''} scenes: Generate 2-4 key metrics (maximum 4)
\end{itemize}

\subsubsection*{\textbf{Output Format (JSON)}}
\begin{lstlisting}[style=jsonstyle]
{
  "scenes": [
    {
      "id": "scene_chart_1",
      "type": "chart",
      "content": {
        "chart_type": "bar_chart",
        "title": "Short, descriptive title",
        "data": [...],
        "data_binding": {
          "x_axis": {"field": "category", "label": "Category"},
          "y_axis": {"field": "value", "label": "Value"}
        },
        "style": {
          "background_color": "{background_color}",
          "container_background": "{container_background}",
          "bar_color": "#5b8ff9",
          "text_color": "#e8eaed",
          "grid_color": "#2a3f5f",
          "axis_color": "#8c98a4"
        },
        "layout": {
          "margin": {"top": 80, "right": 60, "bottom": 100, "left": 100},
          "chart_area": {"width": 1120, "height": 540}
        }
      },
      "insight_summary": "Brief summary of what this chart shows (1-2 sentences)"
    }
  ]
}
\end{lstlisting}

\subsubsection*{\textbf{Critical Rules}}
\begin{itemize}
    \item MUST follow planned\_type (chart or stat\_cards)
    \item MUST use ONLY provided data - all numbers must match exactly, convert wide to long format for bar\_chart
    \item Chart type must match analysis type and have sufficient data points
    \item Include ALL query-requested dimensions/categories (even if zero/small)
    \item Title: clear and descriptive; insight\_summary: describe KEY finding visible in visualization
    \item \textbf{Theme consistency (soft constraint)}: Prefer using the provided theme tokens (e.g., \texttt{background\_color}, \texttt{container\_background}) for a coherent overall look. You MAY introduce subtle per-scene variation (e.g., slight lightness/opacity changes or a gentle gradient) as long as readability and overall theme consistency are preserved.
    \item DO NOT include: ``narration'' field, timing fields (time\_range, time\_start, time\_end)
    \item DO NOT omit dimensions mentioned in query, generate, estimate, or invent data values
\end{itemize}

\subsubsection*{\textbf{Examples}}
\{example\}

\subsubsection*{\textbf{Final Instruction}}
Now generate the visualization configuration. Return ONLY the JSON, nothing else.
\end{examplebox}

\textbf{Example:} The following demonstrates examples of good and bad insight summaries for the Visual Designer Agent. These examples illustrate how to write effective insight summaries that highlight key findings from the visualization.

\subsubsection{Visual Designer Agent Example}
\label{visual-designer-example}
\begin{examplebox}
    {Example Insight Summary}
    \begin{itemize}
        \item Good: ``Q3 sales surged 50\% above Q2, marking the strongest quarterly growth of the year.''
        \item Good: ``San Francisco experiences the highest delay rate at 35 minutes, significantly worse than Boston's 12 minutes.''
        \item Bad: ``This chart shows sales data.'' (too vague)
        \item Bad: ``The analysis indicates variations.'' (not specific)
    \end{itemize}
\end{examplebox}

\subsection{Narrative Director Agent}
\label{appendix:narrative_director_prompt}

\begin{examplebox}{Prompt Template: Narrative Director Agent}
You are a professional video director and scriptwriter specializing in data visualization videos.\\

\textbf{Task}: Given all generated visualization scenes and the chosen \texttt{narrative\_pattern}, arrange them into a dramatically coherent order, write a complete script (opening, scene narrations, closing), and optionally generate a stat\_cards scene that summarizes key metrics from the chart data. 

\subsubsection*{\textbf{Input}}
\begin{itemize}
    \item \textbf{User Query}: \{query\}
    \item \textbf{Narrative Pattern}: \{narrative\_pattern\}
    \item \textbf{Generated Scenes (complete scene configurations)}: \{scenes\_info\}
\end{itemize}

\subsubsection*{\textbf{Narrative Pattern Strategy}}
The \texttt{narrative\_pattern} guides scene ordering, climax placement, pacing, and transition style:
\begin{itemize}
    \item \texttt{freytag\_default}: Setting $\rightarrow$ Rising $\rightarrow$ Climax $\rightarrow$ Resolution; build toward one main finding.
    \item \texttt{hook\_then\_evidence}: lead with the most striking finding, then provide supporting evidence.
    \item \texttt{comparison\_driven}: keep contrasting scenes adjacent and make transitions explicitly signal the contrast.
    \item \texttt{time\_driven}: preserve chronological order and use temporal transitions.
    \item \texttt{drill\_down}: move from overview to segments to drivers, using zoom-in transitions.
\end{itemize}

\subsubsection*{\textbf{Your Tasks}}
\begin{enumerate}
    \item \textbf{Arrange Scene Order (pattern-driven)}: Order scenes by the \texttt{narrative\_pattern} from the Scene Planner. By default (\texttt{freytag\_default}), follow Freytag's Pyramid: Setting (context/overview) $\rightarrow$ Rising (build evidence) $\rightarrow$ Climax (the single most important finding) $\rightarrow$ Resolution (recap), and emit the stage assignment as \texttt{narrative\_arc}. Alternative patterns (\texttt{hook\_then\_evidence}, \texttt{comparison\_driven}, \texttt{time\_driven}, \texttt{drill\_down}) override this default ordering. Stat cards, if generated, belong in the Resolution stage.
    \item \textbf{Generate Opening Scene}: Title (max 8-10 words), Subtitle (5-10 words), Narration (1 sentence, 12-18 words) introducing topic.
    \item \textbf{Write Scene Narrations}: For EACH scene, write 1-2 sentences (10-15 words each), describe chart and highlight key insight, connect with transition phrases for continuous story flow.
    \item \textbf{Generate Stat Cards Scene (Optional)}: When query asks for ``key metrics''/``summary statistics'' or summarizing enhances narrative. Extract metrics DIRECTLY from chart scenes' data, create 2-4 cards (max 4). Prefer the same theme tokens as chart scenes; you MAY apply subtle per-scene background variation while preserving readability and overall theme consistency.
    \item \textbf{Generate Closing Scene}: Title ``Thank You'' or brief message (max 5 words), Narration (1 sentence, 15-20 words) summarizing KEY findings from ALL scenes.
\end{enumerate}

\subsubsection*{\textbf{Output Format (JSON)}}
\begin{lstlisting}[style=jsonstyle]
{
  "narrative_arc": {
    "setting": ["scene_id1"],
    "rising": ["scene_id2"],
    "climax": "scene_id3",
    "resolution": ["scene_stats"]
  },
  "scene_order": ["scene_id1", "scene_id2", "scene_id3"],
  "opening": {
    "id": "scene_opening",
    "type": "opening",
    "content": {
      "title": "Title",
      "subtitle": "Subtitle",
      "style": {
        "text_color": "#ffffff",
        "subtitle_color": "#e0e0e0"
      }
    },
    "narration": [{"text": "Opening narration"}]
  },
  "scene_narrations": {
    "scene_id1": [{"text": "First narration line"}, {"text": "Second line"}]
  },
  "stat_cards": {
    "id": "scene_stats",
    "type": "stat_cards",
    "content": {
      "title": "Key Metrics Summary",
      "cards": [
        {
          "title": "Metric Name",
          "value": "Formatted Value",
          "subtitle": "Unit/Description",
          "color": "#hexcode"
        }
      ],
      "style": {
        "background_color": "#0f1419",
        "container_background": "#0f1419",
        "card_background": "#1a2332",
        "text_color": "#e8eaed",
        "accent_color": "#5b8ff9",
        "border_color": "#2a3f5f"
      }
    },
    "narration": [{"text": "Stat cards narration"}]
  },
  "closing": {
    "id": "scene_closing",
    "type": "closing",
    "content": {
      "title": "Thank You",
      "style": {
        "text_color": "#ffffff",
        "subtitle_color": "#e0e0e0"
      }
    },
    "narration": [{"text": "Closing narration summarizing key findings"}]
  }
}
\end{lstlisting}

\subsubsection*{\textbf{Critical Requirements:}}
\begin{enumerate}
    \item Data and Format Rules
    \begin{itemize}
    \item \textbf{Number Formatting}: $\geq$1M: ``X.X million''/``X.XM'', $\geq$1K: ``X.X thousand''/``X.XK'', $<$1K: exact number, proportions: convert to percentage (1 decimal, e.g., 0.944444 $\rightarrow$ ``94.4\%'')
    \item \textbf{Year/Time Accuracy}: When mentioning specific year/time, MUST use data value for THAT EXACT year, not peak values from other time points
    \item \textbf{Multiple Y-Axis}: If data\_binding.y\_axis is array, mention all relevant metrics in narration
    \item \textbf{Stat Cards Values}: Must match EXACTLY what's shown in chart scenes' data
\end{itemize}

\item Structure and Narration Rules
\begin{itemize}
    \item scene\_order must include ALL scene IDs from input; if generating stat\_cards, add ``scene\_stats'' to scene\_order (typically after chart scenes)
    \item Always emit \texttt{narrative\_arc} (Setting/Rising/Climax/Resolution), assigning every chart scene to exactly one stage
    \item Narration lengths: Opening 1 sentence (12-18 words), Scene 1-2 sentences (10-15 words each), Closing 1 sentence (15-20 words)
    \item Narration must flow as continuous story with transitions, reference specific data/insights from insight\_summary
    \item Use consistent background colors: ``\#0f1419'', ``\#1a2332''
    \item DO NOT include: timing fields (time\_range, time\_start, time\_end, audio\_file), verbose narration, stat\_cards with mismatched values, data from wrong time points
\end{itemize}

\end{enumerate}

\subsubsection*{\textbf{Examples}}
\{example\}

\subsubsection*{\textbf{Final Instruction}}
Now generate the scene order and complete script. Return ONLY the JSON, nothing else.
    
\end{examplebox}

\textbf{Example:} The following demonstrates a concrete example of good narration with smooth transitions between scenes. This example illustrates how to write coherent scene narrations that flow as a continuous story.

\subsubsection{Narrative Director Agent Example}
\label{narrative-director-example}
\begin{examplebox}
    {Good Narration with a Freytag Arc (Setting, Rising, Climax, Resolution)}
    \begin{itemize}
        \item \textit{(Setting)} Scene 1: ``First, the overall landscape. Total revenue reached 4.2 billion across all regions.''
        \item \textit{(Rising)} Scene 2: ``But not all regions grew equally. Europe and the Americas averaged 12 percent.''
        \item \textit{(Climax)} Scene 3: ``Asia surged 58 percent, three times faster than any other region.''
        \item \textit{(Resolution)} Closing: ``Across all regions, growth averaged 18 percent year over year.''
    \end{itemize}
\end{examplebox}

\subsection{Scene Animation Generator Agent}

\begin{examplebox}{Prompt Template: Scene Animation Generator Agent}
You are an animation designer for data videos following the ``narration-animation interplay'' approach.\\

\textbf{Task}: Add entrance and emphasis animations for ONE scene that sync with narration timing.

\textbf{Critical Requirement}: One narration can have multiple emphasis animations - create one animation for EACH entity mentioned in the narration.

\subsubsection*{\textbf{Input}}
\begin{itemize}
    \item \textbf{Current Scene (complete information)}: \{current\_scene\}
    \item \textbf{Context}: \{context\}
    \item \textbf{Position}: \{position\} of \{total\_scenes\} (1-based)
\end{itemize}

\subsubsection*{\textbf{Process}}
\begin{enumerate}
    \item Read narration text carefully, identify ALL entity names (companies, products, categories, data points)
    \item Create ONE entrance animation: bind to trigger\_narration: 0, auto-select effect based on chart\_type
    \item Create ONE emphasis animation for EACH entity: bind to trigger\_narration: $<$index$>$ where entity is first mentioned, use ``pulse'' effect with intensity 0.1, extract data\_filter from entity name
\end{enumerate}

\subsubsection*{\textbf{Animation Effects Guidelines}}
\begin{itemize}
    \item Select entrance and emphasis animations appropriate for the chart type:
    \begin{itemize}
        \item \textbf{Bar/Column Charts}: entrance \texttt{grow\_bars}, \texttt{slide\_in}; emphasis \texttt{pulse}, \texttt{highlight}
        \item \textbf{Line/Area Charts}: entrance \texttt{draw\_line}, \texttt{fade\_in}; emphasis \texttt{pulse}, \texttt{glow}
        \item \textbf{Pie/Donut Charts}: entrance \texttt{grow\_slices}, \texttt{rotate\_in}; emphasis \texttt{pulse}, \texttt{expand}
        \item \textbf{Treemap/Hierarchical}: entrance \texttt{zoom\_in}, \texttt{cascade}; emphasis \texttt{pulse}, \texttt{border\_highlight}
        \item \textbf{Scatter/Bubble Charts}: entrance \texttt{fade\_in}, \texttt{sequential}, \texttt{scale\_in}; emphasis \texttt{pulse}, \texttt{enlarge}
        \item \textbf{Heatmap}: entrance \texttt{fade\_in}, \texttt{cell\_by\_cell}; emphasis \texttt{pulse}, \texttt{intensity\_boost}
        \item \textbf{Stat Cards}: entrance \texttt{fade\_in}, \texttt{slide\_up}; emphasis \texttt{pulse}, \texttt{glow}
        \item \textbf{Opening/Closing}: entrance/exit \texttt{fade\_in/out}, \texttt{slide\_in/out}, \texttt{zoom}
    \end{itemize}
    \item \textbf{Note}: The animation effects above represent commonly used patterns. The system supports custom animation sequences through the declarative synchronization mechanism.
    \item Common emphasis effect: \texttt{pulse} with intensity 0.1 for most chart types
\end{itemize}

\subsubsection*{\textbf{Output Format (JSON)}}
\begin{lstlisting}[style=jsonstyle]
{
  "animations": [
    {
      "id": "entrance_anim",
      "type": "entrance",
      "effect": "grow_bars",
      "trigger_narration": 0,
      "description": "Chart entrance animation"
    },
    {
      "id": "emphasis_entity",
      "type": "emphasis",
      "effect": "pulse",
      "trigger_narration": 1,
      "target_data": {"data_filter": {"field": "value"}},
      "style": {"intensity": 0.1},
      "description": "Highlight entity when mentioned"
    }
  ]
}
\end{lstlisting}

\subsubsection*{\textbf{Critical Rules}}
\begin{itemize}
    \item Include 1 entrance animation (trigger\_narration: 0)
    \item For each entity mentioned in narration, create 1 emphasis animation with correct trigger\_narration index (must not exceed number of narrations)
    \item Do NOT manually set time\_start or duration - use trigger\_narration for timing synchronization
    \item Keep all existing configuration fields unchanged, use consistent animation effects across scenes
\end{itemize}

\subsubsection*{\textbf{Final Instruction}}
Now generate animations for this scene. Return ONLY the JSON with animations array, nothing else.

\end{examplebox}

\subsection{TSX Static Component Agent}

\subsubsection{Static Charts Scene}
\begin{examplebox}{Prompt Template: Static Charts}
    You are creating a CLEAN, MINIMAL infographic for data video narration following ChartGalaxy design principles: SIMPLE, CLEAR, DATA-FOCUSED.

    \vspace{0.8em}
    
    \textbf{CRITICAL CONTEXT:}
    \begin{itemize}
        \item Current Scene: {scene\_index} of {total\_scenes}
        \item A global theme is provided in JSON config (e.g., \texttt{background\_color}, \texttt{container\_background}). Prefer using it to keep the UI coherent across scenes.
        \item You MAY introduce subtle per-scene background variation (small lightness/opacity shifts or gentle gradients) while keeping overall theme consistency and strong text/axis contrast.
        \item Each scene MUST have a UNIQUE visual identity
    \end{itemize}
    
    \textbf{CORE PRINCIPLES:}
    \begin{itemize}
        \item CLEAN \& MINIMAL - No clutter, no excessive decoration
        \item DATA FIRST - Information clarity \textgreater visual effects
        \item SUBTITLE-FRIENDLY - Reserve top 80px for subtitle overlay, bottom 180px for subtitle display
        \item VISUAL CONSISTENCY - Keep typography, spacing, and overall theme coherent across scenes (background may vary subtly)
        \item JSX SAFETY - Escape special characters in JSX text
    \end{itemize}
    
    \textbf{Data to Visualize:}\\
    Title: "{chart\_title}"\\
    Type: {chart\_type}\\
    Canvas: {video\_meta.get('width', 1280)}x{video\_meta.get('height', 720)}px\\
    Data Binding: {json.dumps(data\_binding, indent=2, ensure\_ascii=False)}\\
    
    \textbf{Multiple Y-Axis Support:}
    \begin{itemize}
        \item If y\_axis is an ARRAY $\rightarrow$ Create GROUPED BAR CHART
        \item If y\_axis is a DICT $\rightarrow$ Create standard single bar chart
    \end{itemize}
    
    \textbf{CRITICAL RULES:}
    \begin{itemize}
        \item Use theme tokens from JSON config as defaults (background\_color: {background\_color}, container\_background: {container\_background}). You MAY adjust them slightly per scene (e.g., lightness/opacity/gradient) as long as the overall theme remains consistent and the chart remains highly legible.
        \item Reserve bottom 180px for subtitle display
        \item Match colors to data meaning (gold=money, red=problem, green=growth)
        \item Highlight key data IN THE CHART ITSELF with size/color/stroke
        \item Chart height: $\sim 350-400px$, X-axis labels at $ y \leq 360-370$
        \item Escape JSX special characters (\textless, \textgreater, \&)
        \item DO NOT add redundant information cards or tooltip boxes
        \item DO NOT use the same color scheme for consecutive scenes
    \end{itemize}
    
    \textbf{OUTPUT}: Generate ONLY the complete TypeScript code - NO markdown blocks, NO explanations. 
\end{examplebox}

\subsubsection{Opening, Closing, Stat\_cards Scene}
\begin{examplebox}{Prompt Template: Opening/Closing/Stat Cards Scene}
    You are creating an OPENING/CLOSING/STAT CARDS SCENE for a data video.

    \vspace{0.8em}
    
    \textbf{VIDEO CONTEXT:}
    \begin{itemize}
        \item Video Title: "\{video\_meta.get('title', 'Data Insights')\}"
        \item Scene \{scene\_index\} of \{total\_scenes\}
        \item Duration: \{duration\} seconds
    \end{itemize}
    
    \textbf{SCENE CONTENT:}
    \begin{itemize}
        \item Main Title: "\{title\}"
        \item Subtitle: "\{subtitle\}" (for Opening scene only)
        \item Cards Data: \{cards\_str\} (for Stat Cards scene only)
        \item Narration: "\{narration\_text\}"
    \end{itemize}
    
    \textbf{COMMON DESIGN REQUIREMENTS:}
    \begin{itemize}
        \item \textbf{Background}: Use background color from config (\{bg\_color\})
        \item \textbf{Typography}: Modern sans-serif font (e.g., 'Inter', 'Helvetica', 'Arial')
        \item \textbf{Title}: Large, bold (font-size: 56-72px, font-weight: 700-900), color: \{text\_color\}
        \item \textbf{Subtitle}: Smaller text (font-size: 24-32px, font-weight: 400-500), color: \{subtitle\_color\}
        \item \textbf{NO ANIMATIONS}: All elements at full opacity and final positions (animations added later)
        \item \textbf{Subtitle Space}: Reserve BOTTOM 80-130px for narration subtitles
    \end{itemize}
    
    \textbf{SCENE-SPECIFIC REQUIREMENTS:}
    \begin{itemize}
        \item \textbf{Opening}: Centered title and subtitle, spacing 20-30px between them
        \item \textbf{Closing}: Centered title, optional tagline below (font-size: 20-24px)
        \item \textbf{Stat Cards}: Cards in SINGLE ROW (flexWrap: 'nowrap'), flex: '1 1 0\%', width: 0 on each card. Each card: border 2px, padding 28px 20px, number font-size 52px, label font-size 15px. Reserve bottom 130px for subtitles.
    \end{itemize}
    
    \textbf{OUTPUT}: Generate complete TSX component. See examples: \{Opening\_Example~\ref{tsx-static-example}\}, \{Closing\_Example~\ref{tsx-static-example}\}, \{Stat\_Card\_Example~\ref{tsx-static-example}\}
    
    \textbf{CRITICAL}:
    \begin{itemize}
        \item Replace [ComponentName] with actual component name
        \item Keep structure clean and simple
        \item All styles inline for easy animation later
        \item DO NOT add animation logic
        \item Return ONLY complete TSX code, no explanation
    \end{itemize}
\end{examplebox}

\textbf{Example:} The following demonstrates concrete examples of TSX static component code for different scene types (Opening, Closing, and Stat Cards). These examples illustrate the expected component structure, styling, and layout patterns.

\subsubsection{TSX Static Component Agent Example}
\label{tsx-static-example}
\begin{examplebox}
    {Opening/Closing Scene Example}
    \begin{lstlisting}[style=typescriptstyle]
import React from 'react';
import { AbsoluteFill } from 'remotion';

interface SceneProps {
  sceneStartOffset?: number;
  narrations?: Array<{text: string; time_start: number; time_end: number}>;
}

export const [ComponentName]: React.FC<SceneProps> = ({ sceneStartOffset = 0 }) => {
  return (
    <AbsoluteFill style={{...containerStyle}}>
      <div style={{...titleStyle}}>{title}</div>
      {/* Opening: subtitle with fontSize: 28 */}
      {/* Closing: optional summary text with fontSize: 22, opacity: 0.9 */}
    </AbsoluteFill>
  );
};
    \end{lstlisting}
\end{examplebox}
\begin{examplebox}
    {Stat Card Example}
    \begin{lstlisting}[style=typescriptstyle]
import React from 'react';
import { AbsoluteFill } from 'remotion';

interface StatCard { number: string; label: string; color: string; }
interface SceneProps {
  sceneStartOffset?: number;
  narrations?: Array<{text: string; time_start: number; time_end: number}>;
  title?: string;
}

export const [ComponentName]: React.FC<SceneProps> = ({ 
  sceneStartOffset = 0, 
  title = "{title}" 
}) => {
  const cards: StatCard[] = {cards_json};
  
  return (
    <AbsoluteFill style={{...containerStyle, paddingBottom: 130}}>
      {title && 
        <div style={{...titleContainerStyle}}>
            <h2>{title}</h2>
        </div>}
      <div style={{display: 'flex', flexWrap: 'nowrap', gap: 20, ...}}>
        {cards.map((card, index) => (
          <div key={index} style={{
            flex: '1 1 0%', width: 0, minWidth: 0,
            background: 'rgba(26, 32, 44, 0.8)',
            border: `2px solid ${card.color}`,
            ...cardStyle
          }}>
            <div style={{fontSize: 52, color: card.color, ...}}>{card.number}</div>
            <div style={{fontSize: 15, color: '#e0e0e0', ...}}>{card.label}</div>
          </div>
        ))}
      </div>
    </AbsoluteFill>
  );
};
    \end{lstlisting}
\end{examplebox}

\subsection{Animation Adding Agent}

\subsubsection{Static Charts Animation}
\begin{examplebox}{Prompt Template: Animated Chart Component Specification}
You are a \textbf{React, D3.js, and Remotion animation expert}.

\vspace{0.8em}

\textbf{Task}\\
I have a static D3 infographic component, and I need you to \textbf{only add animation logic} without modifying the static rendering part.

\vspace{0.8em}

\textbf{Scene Information}\\
Title: "\{scene\_title\}"\\
Animation Configuration: \{json.dumps(animations\_config, indent=2)\}\\
Subtitle Configuration: \{json.dumps(narrations, indent=2)\}\\
Existing Static Component Code: \texttt{\{static\_tsx\_code\}}

\vspace{0.8em}

\textbf{What to Do:}
\begin{enumerate}
    \item \textbf{Add Remotion imports}: \texttt{useCurrentFrame}, \texttt{useVideoConfig}
    \item \textbf{Add hooks}: \texttt{const frame = useCurrentFrame(); const \{fps\} = useVideoConfig(); const sceneStartOffset = \{scene\_start\_time\};}
    \item \textbf{Add second useEffect} for animations:
    \begin{itemize}
        \item Implement entrance and emphasis animations
        \item After entrance ends, force all elements to final state
        \item Use precise class selectors (\texttt{.bar}, \texttt{.dot}, etc.), not broad tag selectors
        \item Always check if data \texttt{d} exists before accessing properties
        \item Use \texttt{.style('opacity', ...)} consistently, not mixed with \texttt{.attr()}
    \end{itemize}
    \item \textbf{Add subtitle display}: Create \texttt{getCurrentNarration()} helper, display in bottom area (reserved 35-130px)
    \item \textbf{Modify initial state}: Set initial animation values (e.g., height: 0, opacity: 0) in first useEffect
\end{enumerate}

\vspace{0.8em}

\textbf{Key Requirements:}
\begin{itemize}
    \item Use absolute time (seconds) for animation timing, not relative progress
    \item Handle multiple simultaneous animations correctly (use Set to track highlighted items)
    \item Use standard D3 ease functions (e.g., \texttt{d3.easeCubicOut(progress)})
    \item DO NOT modify core static rendering logic or JSX layout structure
\end{itemize}

\vspace{0.8em}

\textbf{Output}: Complete TypeScript code with animations. Component name: \texttt{export const SceneComponentAnimated}. No markdown blocks, no explanations.
\end{examplebox}

\subsubsection{Opening, Closing, Stat\_cards Scene Animation}
\begin{examplebox}{Prompt Template: Opening/Closing/Stat Cards Scene Animation}
You are adding \textbf{ANIMATIONS} to an \textbf{OPENING/CLOSING/STAT CARDS SCENE} component.

\vspace{0.8em}

\textbf{SCENE TIMING:}
\begin{itemize}
    \item Scene starts at: \{scene\_start\_time\}s
    \item Duration: \{duration\}s
    \item Number of cards: \{num\_cards\} (for Stat Cards only)
    \item FPS: 30
\end{itemize}

\vspace{0.8em}

\textbf{ANIMATION CONFIGURATION:} (for Stat Cards only)\\
\{animations\_json\}

\vspace{0.8em}

\textbf{COMMON REQUIREMENTS:}
\begin{itemize}
    \item \textbf{CRITICAL}: \texttt{frame} in Sequence is LOCAL (starts from 0), NOT global frame number
    \item Use \texttt{relativeTime = frame / fps} for animation timing
    \item Use \texttt{absoluteTime = sceneStartOffset + relativeTime} for subtitle matching
    \item Import Remotion hooks: \texttt{useCurrentFrame}, \texttt{useVideoConfig}, \texttt{interpolate}, \texttt{Easing}
    \item For Stat Cards: Also import \texttt{useMemo}, add \texttt{animations?: Animation[]} to SceneProps
    \item Apply animations to individual elements separately (title, subtitle, cards), NOT to AbsoluteFill
    \item Use \texttt{interpolate} with \texttt{Easing.out(Easing.cubic)} for smooth animations
    \item Add narration subtitle UI at bottom
    \item DO NOT change component structure or content
\end{itemize}

\vspace{0.8em}

\textbf{SCENE-SPECIFIC ANIMATIONS:}
\begin{itemize}
    \item \textbf{Opening}: Title and subtitle fade in with slide-up effect
    \item \textbf{Closing}: Title fade in + slide up (30px) + scale (0.92$\rightarrow$1.0) over 0.8s, starts at 0.15s. Subtitle fade in + slide up (20px) over 0.7s, starts at 0.5s. NO fade out.
    \item \textbf{Stat Cards}: Cards appear sequentially using config's \texttt{stagger\_delay}. Each card: fades in + scales up + slides up. Emphasis: border width 3px + glowing shadow + pulse (2.5 cycles) when active. MUST use \texttt{animations} prop from config.
\end{itemize}

\vspace{0.8em}

\textbf{Examples}: \{Opening\_Example~\ref{animation-adding-example}\}, \{Closing\_Example~\ref{animation-adding-example}\}, \{Stat\_Card\_Example~\ref{animation-adding-example}\}

\vspace{0.8em}

\textbf{OUTPUT}: Return ONLY the complete animated TSX code, no explanation.
\end{examplebox}

\textbf{Example:} The following demonstrates concrete examples of animation logic for different scene types (Opening, Closing, and Stat Cards). These examples illustrate how to add Remotion animation hooks and implement entrance/emphasis animations that sync with narration timing.

\subsubsection{Animation Adding Agent Example}
\label{animation-adding-example}
\begin{examplebox}
    {Opening/Closing Scene Animation Example}
    \begin{lstlisting}[style=typescriptstyle]
export const ComponentName: React.FC<SceneProps> = ({ 
  sceneStartOffset = 0, 
  narrations = [] 
}) => {
  const frame = useCurrentFrame();
  const { fps } = useVideoConfig();
  const relativeTime = frame / fps;
  const absoluteTime = sceneStartOffset + relativeTime;
  
  // Opening: Simple fade + slide up
  const titleProgress = Math.max(0, Math.min(1, (relativeTime - 0.2) / 0.6));
  const titleOpacity = titleProgress;
  const titleY = (1 - titleProgress) * 20;
  
  // Closing: Rich layered animation with interpolate
  // const titleProgress = interpolate(frame, [titleDelay * fps, ...], [0, 1], 
  //   { easing: Easing.out(Easing.cubic) });
  // const titleScale = 0.92 + 0.08 * titleProgress;
  
  const currentNarration = narrations.find(
    n => absoluteTime >= n.time_start && absoluteTime < n.time_end
  );
  
  return (
    <AbsoluteFill>
      <div style={{...originalTitleStyle, opacity: titleOpacity, 
        transform: `translateY(${titleY}px)`}}>
        {title}
      </div>
      {currentNarration && (
        <div style={{position: 'absolute', bottom: 35, ...subtitleStyle}}>
          {currentNarration.text}
        </div>
      )}
    </AbsoluteFill>
  );
};
    \end{lstlisting}
\end{examplebox}
\begin{examplebox}
    {Stat Card Animation Example}
    \begin{lstlisting}[style=typescriptstyle]
export const ComponentName: React.FC<SceneProps> = ({ 
  sceneStartOffset = 0,
  narrations = [],
  animations = []  // MUST receive animations prop
}) => {
  const frame = useCurrentFrame();
  const { fps } = useVideoConfig();
  const relativeTime = frame / fps;
  const absoluteTime = sceneStartOffset + relativeTime;
  
  // Extract animations from config
  const entranceAnim = useMemo(() => 
    animations.find(a => a.type === 'entrance'), [animations]);
  const emphasisAnims = useMemo(() => 
    animations.filter(a => a.type === 'emphasis'), [animations]);
  
  // Card entrance: sequential appearance with stagger
  const getCardProgress = (index: number) => {
    const cardStartFrame = entranceStartFrame + index * (staggerDelay * fps);
    return interpolate(frame, [cardStartFrame, cardStartFrame + duration], 
      [0, 1], { easing: Easing.out(Easing.cubic) });
  };
  
  // Card emphasis: pulse effect when mentioned in narration
  const getCardEmphasis = (cardIndex: number) => {
    const emphasisAnim = emphasisAnims.find(a => 
      a.target_data?.card_index === cardIndex);
    if (!emphasisAnim || !isActive) return { scale: 1, borderWidth: 2 };
    const pulse = Math.sin(progress * Math.PI * 2.5) * intensity * 0.4 + 1;
    return { scale: pulse, borderWidth: 3, 
      boxShadow: `0 0 12px ${card.color}40, 0 0 24px ${card.color}20` };
  };
  
  return (
    <AbsoluteFill>
      {cards.map((card, index) => {
        const progress = getCardProgress(index);
        const emphasis = getCardEmphasis(index);
        return (
          <div key={index} style={{
            ...originalCardStyle,
            opacity: progress,
            transform: `scale(${0.8 + 0.2 * progress} * ${emphasis.scale}) 
              translateY(${(1 - progress) * 30}px)`,
            border: `${emphasis.borderWidth}px solid ${card.color}`,
            boxShadow: emphasis.boxShadow
          }}>
            {/* Card content */}
          </div>
        );
      })}
    </AbsoluteFill>
  );
};
    \end{lstlisting}
\end{examplebox}

\section{Evaluation Framework and Criteria}

% Control table placement and page breaks
\setlength{\floatsep}{12pt plus 2pt minus 2pt}
\setlength{\textfloatsep}{20pt plus 2pt minus 4pt}
\setlength{\intextsep}{12pt plus 2pt minus 2pt}

\subsection{Detailed Evaluation Framework and Scoring Criteria}
\label{appendix:detailed_evaluation_questions}

This appendix provides a multi-dimensional assessment system for the generated videos. The design of this system takes into account the \textbf{User Intent }as well as the three core components of the data videos: \textbf{Visualization}, \textbf{Narration}, and \textbf{Animation}. We have defined the following five core indicators (F1-F5), and have formulated detailed 1-5 point scoring standards.

\noindent\textbf{F1: Intent Fulfillment} 
This perspective evaluates whether the data video solved the user's problem, rather than just whether it was made correctly. It focuses on the relevance, completeness, and clarity of the answer provided.

\begin{examplebox}{Prompt: Intent Fulfillment Description and Scoring Criteria (1-5 Scale)}
% \textbf{Data Fidelity:}

% \vspace{0.5em}
\textbf{Description:} \\
The primary goal of a data video is to answer the user's analytical question. Intent Fulfillment evaluates whether the video answers the user's query, including the completeness of the query response, the achievement of analytical goals, and content relevance. This dimension focuses not on "whether the visualization is well done," but on "whether it solves the user's problem."

\vspace{1em}
\textbf{Scoring criteria(Strict):}
\begin{itemize}
  \item[\textbf{S 1.}] Completely fails to answer the user's questions, insights are irrelevant to user intent, content deviates from the topic.  
  
  \item[\textbf{S 2.}] Only surface questions are answered, no deep analysis, insights are weakly connected to user intent, style is mismatched. Or, severe misalignment significantly hinders understanding.\textbf{ This score should only be given when the problem is so severe that the video becomes completely incomprehensible.}
  
  \item[\textbf{S 3.}] Some questions are answered, insights are partially relevant, style is acceptable. Or, alignment issues cause moderate confusion that hinders understanding.
  
  \item[\textbf{S 4.}]  Main questions are answered, 1-2 minor points may be missed, insights are relevant but could be deeper, style is appropriate, no obvious visual-narrative misalignment.
  
  \item[\textbf{S 5.}] Only when all user questions are fully answered, the insights are highly relevant to the user's intent, and the video style perfectly matches the target audience. This is exceptional and rare. \textbf{Evaluation Focus: Video presentation quality (smooth animation, clear narration, visual completeness), not data accuracy.}
\end{itemize}
\end{examplebox}

\vspace{1em}
\noindent\textbf{F2: Data Insight} 
This viewpoint stresses that data insights must be visually proven to be credible. The value of an insight depends on whether it can be clearly seen and verified in the visualization.
\begin{examplebox}{Prompt: Data Insight Description and Scoring Criteria (1-5 Scale)}
% \textbf{Data Fidelity:}

% \vspace{0.5em}
\textbf{Description:} \\
Data Insight evaluates the quality of discoveries from the data, including the depth, novelty, relevance, and value of insights. This dimension focuses on meaningful findings and patterns extracted from the data, and whether these insights help users understand the story behind the data. Insights must be visually verifiable—if an insight is only verbally stated and cannot be verified or understood from visual elements, its credibility and value are diminished.

\vspace{1em}
\textbf{Scoring criteria(Strict):}
\begin{itemize}
  \item[\textbf{S 1.}] No meaningful insights are presented, only raw data display without analysis, or insights are completely irrelevant.
  
  \item[\textbf{S 2.}] Insights are superficial, only reveal obvious patterns, provide limited value. Or, the visuals fail to demonstrate the stated insights, making them unverifiable. \textbf{If insights are not adequately visually supported (e.g., narration describes dynamic comparisons but visuals are static, or insights cannot be verified from the visuals), a score of 2 should be given.}
  
  \item[\textbf{S 3.}] Insights are acceptable but relatively shallow, reveal basic patterns, provide some information but lack depth. Insights have some visual support, but may not be clear or complete.
  
  \item[\textbf{S 4.}]  Insights are meaningful and relevant, reveal some patterns, provide useful information, insights are visually supported but could be deeper.
  
  \item[\textbf{S 5.}] Only when insights are extremely profound and meaningful, reveal novel patterns, are highly relevant to the user query, provide significant value, and are clearly showcased visually. This requires exceptional analytical depth. \textbf{Evaluation Focus: Quality of insight presentation (whether clearly presented visually), not data accuracy.}
\end{itemize}
\end{examplebox}

\vspace{1em}
\noindent\textbf{F3: Narrative Quality} 
This viewpoint judges how well a story is structured and how clearly it connects to the data insights. It focuses on the storytelling itself, not on whether the data is accurate.

\begin{examplebox}{Prompt: Narrative Quality Description and Scoring Criteria (1-5 Scale)}
% \textbf{Data Fidelity:}

% \vspace{0.5em}
\textbf{Description:} \\
Narrative Quality evaluates the logical quality of the video's narration, including narrative coherence, logical completeness, structural soundness, and the connection between the narrative and data insights. This dimension focuses on the overall logical flow of the story, including the completeness of the beginning, development, climax, and conclusion, as well as natural transitions between scenes.

\vspace{1em}
\textbf{Scoring criteria(Strict):}
\begin{itemize}
  \item[\textbf{S 1.}] Narration is chaotic and disorganized, the storyline is incomprehensible, insights are missing or severely deviated, and there is no clear connection between scenes.

  \item[\textbf{S 2.}]Narration is insufficiently coherent or has serious logical gaps, story structure is incomplete or has logical jumps, scene transitions are abrupt, making the narration difficult to follow. \textbf{If there is a serious disconnect between the narration and the visual presentation (e.g., narration describes a dynamic process but visuals are static), a score of 2 should be given.}
  
  \item[\textbf{S 3.}] Narration is basically coherent but relatively plain, story structure is basically complete but lacks highlights, insight connection is basically clear but not deep enough, scene transitions are acceptable. Minor inconsistencies between narration and visuals (e.g., numerical differences due to rounding) are tolerable.
  
  \item[\textbf{S 4.}] Narration is coherent and logically clear, story structure is basically complete, insights are clearly connected, narration and visuals are consistent, scene transitions are reasonable.
  
  \item[\textbf{S 5.}] Only when the narration is extremely coherent and logically clear, the story structure is complete and well-organized, perfectly connected to insights, the narration and visuals are completely consistent, scene transitions are natural and smooth, achieving a professional narrative level. \textbf{Evaluation Focus: Fluency, structure, and logic of the narration (whether scene transitions are natural, narrative structure is clear, story flow is coherent), not data accuracy.}
\end{itemize}
\end{examplebox}

\vspace{1em}
\noindent\textbf{F4: Animation Effectiveness} 
This aspect assesses how precisely animations are synchronized with the narration and visuals to highlight key elements at the right moments. 
\begin{examplebox}{Prompt: Animation Effectiveness Description and Scoring Criteria (1-5 Scale)}
% \textbf{Data Fidelity:}

% \vspace{0.5em}
\textbf{Description:} \\
Evaluates the synchronization of animations with the narration: whether visual elements (highlights, appearances, annotations, etc.) emphasize the content being discussed at the appropriate moment. This dimension only focuses on animation timing and pointing accuracy, not data or chart quality. Score based on the entire video; only downgrade significantly when problems are recurring or severely impair understanding.

\vspace{1em}
\textbf{Scoring criteria(Strict):}
\begin{itemize}
  \item[\textbf{S 1.}]  Frequent and severe misalignments/highlighting errors, severely interfering with understanding or making it almost impossible to understand.

  \item[\textbf{S 2.}]  Multiple noticeable misalignments (\textgreater 1 second) or frequently highlights wrong elements/points to wrong categories, causing viewers significant difficulty in matching narration with visuals, and significantly hindering understanding.
  
  \item[\textbf{S 3.}] Overall can follow along (\textless 1 second), but timing is unstable, omissions are frequent, or highlighting doesn't closely "follow the narration"; can still basically understand the main points.

  \item[\textbf{S 4.}] Generally well synchronized (<0.5 seconds), highlights correct elements most of the time, occasional minor timing issues or omissions.

  \item[\textbf{S 5.}] Almost perfectly synchronized with narration (<0.2 seconds), highlights/appears correct elements every time, with accurate timing.

\end{itemize}
\end{examplebox}

\vspace{1em}
\noindent\textbf{F5: Aesthetic Quality} 
This point evaluates the visual polish and design coherence of a video, focusing on its color scheme, typography, and layout. It judges whether the aesthetic presentation appears professionally finished and credible, or rough and amateurish.

\begin{examplebox}{Prompt: Aesthetic Quality Description and Scoring Criteria (1-5 Scale)}
% \textbf{Data Fidelity:}

% \vspace{0.5em}
\textbf{Description:} \\
Evaluates the overall visual aesthetics and professionalism of the video, including color scheme coordination, typography refinement (margins, alignment, whitespace), design details (rounded corners, shadows, transition effects), and style consistency. Aesthetic quality reflects the video's "polish." High-quality videos feel professional and carefully crafted; low-quality videos appear rough and amateurish.

\vspace{1em}
\textbf{Scoring criteria(Strict):}
\begin{itemize}
  \item[\textbf{S 1.}] Color scheme is chaotic and jarring, typography has serious problems, completely lacks design sense, style is completely inconsistent, giving the impression of being haphazardly assembled.
  
  \item[\textbf{S 2.}] Multiple design issues affecting visual appeal or readability (obvious overlap, severe cropping, chaotic alignment, placeholder feel, etc.) making core information hard to read or significantly reducing professionalism.
  
  \item[\textbf{S 3.}] Overall readable and basically neat, but details are average; minor local flaws (e.g., slight overlap/cropping/small font size) are allowed as long as they do not affect core information acquisition and are not recurring problems. 
  
  \item[\textbf{S 4.}]  Color scheme and typography are clear and unified, details are reasonably executed, overall appearance is professional; basically no obvious flaws like cropping/overlap.
  
  \item[\textbf{S 5.}] Only when it reaches a professional designer's level: color scheme is refined and harmonious, typography is polished and meticulous, details are well-crafted, overall style is highly consistent, visuals are complete and correctly rendered, with strong visual appeal. \textbf{Evaluation Focus: Visual presentation (color scheme, typography, design details, visual completeness), not data accuracy.}
\end{itemize}
\end{examplebox}

\vspace{1em}

\subsection{MLLM-as-a-Judge Prompt}
To achieve automatic evaluation, we have designed the following prompt to guide the scoring of Gemini-2.5-Pro. \\
The evaluation prompt involved the following components:
\begin{itemize}
    \item \texttt{\{user\_query\}}: User's original intension or queries of the data.
    
    \item \texttt{\{F1\_prompt\}}: Detailed prompt of User Intent in Appendix ~\ref{appendix:detailed_evaluation_questions}

    \item \texttt{\{F2\_prompt\}}: Detailed prompt of Data Insight in Appendix ~\ref{appendix:detailed_evaluation_questions}
    
    \item \texttt{\{F3\_prompt\}}:Detailed prompt of Narrative Quality in Appendix ~\ref{appendix:detailed_evaluation_questions}
    
    \item \texttt{\{F4\_prompt\}}:Detailed prompt of Animation Effectiveness in Appendix ~\ref{appendix:detailed_evaluation_questions}
    
    \item \texttt{\{F5\_prompt\}}:Detailed prompt of Aesthetic Quality in Appendix ~\ref{appendix:detailed_evaluation_questions}
    
\end{itemize}
\begin{examplebox}{Prompt Template: Data Video Evaluation}

You are a professional data visualization video evaluation expert. Please watch this data visualization video carefully and provide a comprehensive assessment.

\vspace{0.8em}
\textbf{Step 1 Video Content Description}:\\
First, provide a detailed description of what you actually observe in the video. The description should include:
\begin{itemize}
    \item The data being visualized (topic, metrics, time period, etc.)
    \item The types of charts/visualizations displayed
    \item The narrative content presented (what the narration says, what text appears on screen)
    \item The overall structure and flow of the video (describe only what you actually see, do not assume)
    \item Key scenes or segments and their sequence
    \item Any major issues hindering comprehension (e.g., contradictions between narration and visuals, missing promised animations, placeholder/incomplete visual elements, cropped/truncated charts, overlapping elements, misaligned layout)
\end{itemize}
\textbf{Key Requirement}: Describe only what you truly see and hear. Do not fabricate, speculate, or assume content not explicitly present. Remain objective and factual.
\\
At the end of the description, include a concise list of Observed Issues (if any), strictly based on what you saw/heard.

\vspace{0.8em}

\texttt{\{user\_query\}}

\vspace{0.8em}
\textbf{Step 2: Evaluation Dimensions and Scoring Criteria}

After providing the video content description, please provide a score of 1-5 (with 1 being the lowest and 5 being the highest) for each of the following five dimensions:

\vspace{0.8em}

F1:User Intent: \texttt{\{F1\_prompt\}}

\vspace{0.8em}

F2:Data Insight: \texttt{\{F2\_prompt\}}

\vspace{0.8em}

F3:Narrative Quality: \texttt{\{F3\_prompt\}}

\vspace{0.8em}

F4:Animation Effectiveness: \texttt{\{F4\_prompt\}}

\vspace{0.8em}

F5:Aesthetic Quality: \texttt{\{F5\_prompt\}}

\vspace{0.8em}

\textbf{Scoring Requirements:}\\
Please provide a 1-5 score for each dimension according to the above scoring criteria, accompanied by a brief justification for the score (1-2 sentences explaining the specific performance for each dimension).\\
For each dimension, the justification must cite at least one specific, observable piece of evidence from the video (e.g., a specific on-screen element, a missing/cropped chart, overlapping elements, a described but missing animation, or a specific mismatch).\textbf{ Do not award a high score based solely on narrative claims if the visual elements fail to show or support them.}

\vspace{0.8em}

\textbf{Key Evaluation Principles:}
\begin{itemize}
    \item Evaluate based solely on what you actually observe in the video. Do not fabricate, speculate, or assume structures, patterns, or content that are not clearly visible or audible in the video.
    \item When describing narrative structure or any other aspect, mention only what you truly see and hear. If you cannot clearly identify something, do not mention it or use vague language like "there seems to be" or "appears to follow".
    \item Be accurate and honest. It is better to give a lower score based on actual observation than a higher score based on assumptions or fabricated structure.
    \item Consider overall quality comprehensively; do not penalize heavily for minor errors: The evaluation should consider the video's overall quality comprehensively (animation smoothness, aesthetic quality, narrative clarity, visual completeness, etc.), rather than assigning a very low score directly due to a localized issue (e.g., data contradiction, minor inconsistency). Significant score reduction should only occur if a problem is severe enough to impact the overall viewing experience or render the video completely incomprehensible.
    \item Distinguish Major Issues from Minor Discrepancies: Only significant inconsistencies (e.g., order-of-magnitude differences, opposite trends, incorrect categories, missing key elements) should be heavily penalized. Minor numerical differences (e.g., due to rounding or unit conversion) are acceptable and should not significantly lower the score.\textbf{For Aesthetic Quality (F5)}: A score of 2 or below is only for issues that seriously harm readability and professionalism. Minor aesthetic flaws not significantly impairing the overall experience are acceptable and merit a score of 3.
\textbf{For Animation Effectiveness (F4)}: Distinguish between animation issues (synchronization, timing, highlight accuracy) and data issues (missing data, incomplete charts). Inability to highlight due to missing data is not an animation misalignment and should impact F1/F2/F3, not F4.
    \item Interlinked Dimensions: A major issue in one dimension often affects others. For example, severe animation-narration misalignment (low F4) can hinder comprehension (affecting F1) and disrupt narrative flow (affecting F3). Do not evaluate dimensions in isolation; consider how problems in one area impact the overall viewing experience.
    
\end{itemize}

Please return the results following the json format below:

\vspace{0.8em}

\begin{lstlisting}[style=jsonstyle]
    {
    "video_description": "A detailed and objective description of the content actually observed in the video, including data themes, chart types, narrative content, structure, and key scenes. Include only what is truly seen and heard, without speculation or assumptions.",
    "scores": {
        "intent_fulfillment": {
            "score": 4,
            "reason": "Addresses the main user questions, insights are relevant to the intent, video style is suitable for the target audience."
        },
        "data_insights": {
            "score": 4,
            "reason": "Insights are meaningful and relevant, revealing some patterns, providing useful information, but could be deeper."
        },
        "narrative_quality": {
            "score": 4,
            "reason": "The narrative is coherent and logically clear, the story structure is basically complete, insights are clearly connected, and scene transitions are reasonable."
        },
        "animation_effectiveness": {
            "score": 4,
            "reason": "Animation synchronization is good (deviation <0.5 seconds), mostly highlights the correct elements, alignment is generally accurate."
        },
        "aesthetic_quality": {
            "score": 4,
            "reason": "The color scheme is harmonious and comfortable, typography is clear and reasonable, with some attention to detail, overall style is basically unified."
        }
    },
    "overall_score": 4.33,
    "summary": "Overall quality is excellent, with outstanding performance in data fidelity and visualization effectiveness, good animation synchronization, and an engaging narrative. Suggestion: Further optimize intent fulfillment to ensure all user questions are fully addressed."
}
\end{lstlisting}

\vspace{0.8em}

Please ensure that the returned content is in valid JSON format and does not contain any markdown code block tags or anything other than JSON

\end{examplebox}

% --- Appendix E: Expert Validation Details ---
\section{Details of Human Expert Validation and Evaluation Interface}
\label{appendix:expert_validation}

\begin{figure*}[t!]
    \centering
    \includegraphics[width=0.77\linewidth]{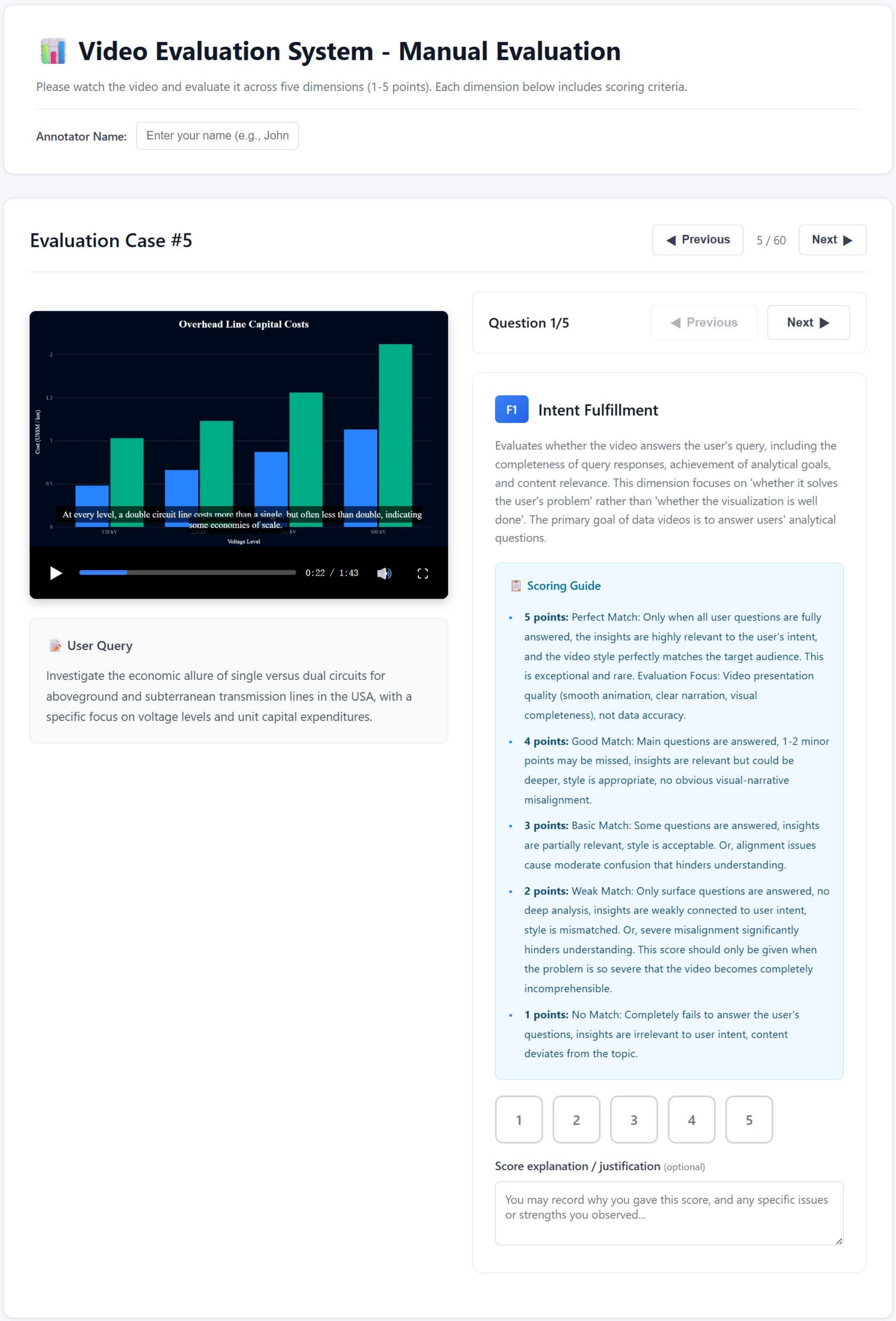}
    \caption{The online evaluation platform used for human expert validation. The interface integrates the original user query, synchronized video playback, a multi-dimensional scoring guide, and a feedback module for documenting qualitative observations.}
    \label{fig:human_eval_system}
\end{figure*}

To verify the effectiveness of the MLLM-based automated evaluation method, we conducted a human expert consistency validation experiment to establish a high-quality ground-truth baseline. The evaluation panel consisted of three experts with professional backgrounds in data analysis and video production. 

To ensure broad representation and coverage of the complete quality spectrum, we employed a dual-layered stratified sampling strategy. First, we performed a preliminary screening from a sample pool generated by various methods (Direct vs. \system) and base models (GPT-5, Claude-Sonnet-4, Gemini-2.5-Pro, DeepSeek-V3.2). Subsequently, we sampled 15 video cases for each of the four quality intervals: 1-2, 2-3, 3-4, and 4-5 points. Through this balanced mechanism across both models and scores, we constructed a validation set of 60 videos, supporting broad coverage of the quality distribution for expert validation.

\begin{table}[H]
  \caption{Automated Evaluator--Human Expert Consistency Metrics.}
  % \vspace{-0.5em}
  \label{tab:consistency_metrics}
  \centering
  \begin{scriptsize}
    %\begin{sc}
    \setlength{\tabcolsep}{4pt}
      \begin{tabular}{lccc}
        \toprule
        \textbf{Dimension} & \textbf{Pearson $r$} & \textbf{MAE} & \textbf{MSE} \\
        \midrule
        Overall & 0.91 & 0.38 & 0.23 \\
        Intent & 0.78 & 0.57 & 0.49 \\
        Insight & 0.76 & 0.51 & 0.47 \\
        Narrative & 0.83 & 0.60 & 0.59 \\
        Animation & 0.90 & 0.47 & 0.48 \\
        Aesthetic & 0.77 & 0.73 & 0.84 \\
        \bottomrule
      \end{tabular}
    %\end{sc}
  \end{scriptsize}
  % \vspace{-1.5em}
\end{table}

The expert review process was conducted via the online evaluation platform provided for this study. As illustrated in Figure~\ref{fig:human_eval_system}, the interface defines a standardized audit workflow for experts: they first examine the original User Query module on the lower-left to identify the analytical objectives the video is expected to achieve; then, they use the interactive player for in-depth observation, focusing on the precision of audio-visual synchronization and narrative coherence. During the scoring phase, experts refer to the detailed Scoring Guide provided in Appendix~\ref{appendix:detailed_evaluation_questions} to provide quantitative ratings (1-5 scale) across five dimensions: Intent Fulfillment, Data Insight, Narrative Quality, Animation Effectiveness, and Aesthetic Quality.

In practice, experts spent approximately 15 to 30 minutes reviewing each video case. This extended period of meticulous review ensured that experts could identify subtle logical deviations, cross-modal numerical conflicts, or temporal misalignments. In addition to quantitative scores, experts utilized the feedback area at the bottom of the interface to document qualitative defects.

\textbf{Consistency Results.} Table~\ref{tab:consistency_metrics} summarizes the agreement between the automated evaluator and expert scores. Overall scores exhibit a strong positive correlation (Pearson $r = 0.91$, $p < 0.001$) with a mean absolute error of 0.38. All dimension-level correlations exceed 0.75, with the strongest agreement for Animation ($r = 0.90$), followed by Narrative ($r = 0.83$); Intent, Aesthetic, and Insight achieve correlations of 0.78, 0.77, and 0.76, respectively. Although Aesthetic has the largest absolute error ($MAE = 0.73$), the results indicate that the automated evaluator largely preserves experts' relative quality judgments across dimensions and quality levels.

\FloatBarrier
\bibliographystyle{abbrv-doi-hyperref}
\bibliography{reference}

\else

%% ---- Revision letter now builds separately via revision/revision_report.tex ----
%% (kept out of the paper; uncomment only if you want a combined letter+paper PDF)
% \pagestyle{plain}
% \pagenumbering{roman}
% \input{revision/revision-letter}
% \clearpage
% \pagenumbering{arabic}
%% -----------------------------------------------------------------

%%%%%%%%%%%%%%%%%%%%%%%%%%%%%%%%%%%%%%%%%%%%%%%%%%%%%%%%%%%%%%%%
%%%%%%%%%%%%%%%%%%%%%% START OF THE PAPER %%%%%%%%%%%%%%%%%%%%%%
%%%%%%%%%%%%%%%%%%%%%%%%%%%%%%%%%%%%%%%%%%%%%%%%%%%%%%%%%%%%%%%%

\maketitle

\section{Introduction}

Data videos combine dynamic charts, voice narration, and animation effects into coherent temporal narratives, and have been widely adopted in business reporting, journalism, and education~\cite{shen2025reflecting,amini2015understanding,shen2022towards,li2026deepeye}.
Compared with static charts or dashboards, data videos guide audiences along a curated narrative path to understand data, offering significant advantages in communication efficiency and audience engagement~\cite{amini2015understanding}.
However, producing a high-quality data video requires multidisciplinary expertise spanning data analysis, narrative design, and video editing, resulting in high production costs and long cycles that severely limit the scalability of this medium.

Existing approaches simplify parts of this workflow but none achieves end-to-end automation from raw data to a complete video.
Static visualization tools (e.g., DeepEye~\cite{luo2018deepeye,luo2020steerable}, HAIChart~\cite{xie2024haichart}, DeepVIS~\cite{shuai2025deepvis}) automate chart generation but lack narrative logic and animation, producing isolated static graphics.
Authoring tools (e.g., WonderFlow~\cite{wang2025wonderflow}, Data Playwright~\cite{shen2025dataplaywright}) introduce narrative-centered paradigms that allow users to add animations and narration to individual charts, but they take pre-made visualizations as their starting point, cannot extract insights from raw data, and do not support multi-scene narrative orchestration.
Pixel-level video generation models (e.g., Sora~\cite{liu2024sora}, Veo3~\cite{deepmind2025veo3}) can generate videos end-to-end, but their black-box nature frequently produces numerical hallucinations and cannot trace visual elements back to underlying data records.
In summary, existing approaches struggle to simultaneously achieve data fidelity, narrative coherence, and end-to-end automation.

Our key observation is that effective data videos are fundamentally structured narratives rather than simple assemblies of visual elements~\cite{shen2025reflecting}.
A well-crafted data video typically consists of multiple scenes, each organized around a distinct analytical insight with coordinated charts, narration, and animations; scenes follow narrative logic such as macro-to-micro or phenomenon-before-cause.
This inspires us to model end-to-end generation as a hierarchical content orchestration problem: starting from a user query, the system makes decisions at the level of macro narrative structure, scene-level content, and micro animation details.
This modeling introduces two core challenges: (1)~how to design a unified intermediate representation that precisely describes charts, narration, animations, and their temporal relationships while ensuring data provenance; (2)~how to efficiently search the vast design space of insights, chart types, and narrative paths for globally coherent compositions across multiple scenes.

To address these challenges, we propose \system, which authors data videos through \textit{declarative multi-agent orchestration} that combines a declarative specification for unified audio-visual representation with a multi-agent pipeline for narrative-coherent scene generation.
For challenge~(1), we design \grammar (Data Video Specification), a declarative specification that decomposes data videos into an ordered scene sequence, where each scene contains content, narration, and animation components.
\grammar binds visual and animation elements to underlying data fields through data-driven semantic references, and replaces manual timestamps with narration-indexed triggering to achieve declarative audio-visual synchronization (Section~\ref{sec:grammar}).
For challenge~(2), we propose a ``Generate-then-Orchestrate'' multi-agent strategy: in the generation stage, agents with distinct roles collaborate in parallel to produce diverse candidate scenes; in the orchestration stage, scenes are globally selected and ordered based on insight value and query coverage, with context-aware narration generated to ensure narrative coherence.
\grammar further serves as a shared editable state supporting three interaction modes (canvas manipulation, script editing, and natural language commands), bridging full automation with fine-grained human control (Section~\ref{sec:system_framework}).

We evaluate \system on 109 real-world samples across five quality dimensions and conduct a within-subjects user study.
Results show that even the most advanced LLM (GPT-5) achieves only 2.13/5 with execution success rates between 48.62\% and 86.24\%; \system improves quality to 3.89 (+83\%) with success rates above 95\%.
The user study further confirms that, compared to a conversational LLM workflow, \system significantly improves creation efficiency (79.7\% reduction in task time) and reduces perceived cognitive load.

The main contributions of this paper include:
(1)~\textbf{\grammar}: a declarative specification for data videos that unifies charts, narration, and animations with their temporal relationships through semantic references and narration-indexed triggering;
(2)~\textbf{Multi-agent framework}: a ``Generate-then-Orchestrate'' two-stage strategy for parallel candidate scene generation and global narrative orchestration;
(3)~\textbf{Interactive system}: a complete web-based system supporting three complementary interaction modes over a shared \grammar state; and
(4)~\textbf{Comprehensive evaluation}: systematic experiments on 109 real-world samples and a within-subjects user study ($N=12$) confirming practical utility.

\section{Related Work}

\subsection{Data Video Authoring}
Prior research has established theoretical foundations for data video creation from perspectives including narrative structure~\cite{amini2015understanding}, animation design primitives~\cite{thompson2020understanding}, and motion design space~\cite{shi2021communicating}, among others~\cite{yang2022design}.
Building on these foundations, various authoring tools have been developed to lower production barriers.
\rev{Guided by Chen et al.'s automation-level taxonomy of narrative visualization tools~\cite{chen2024automation}, we focus on two categories most relevant to data-video creation: human-in-the-loop authoring and automated generation.}

\rev{\emph{Human-in-the-loop authoring and animation-design approaches} retain substantial user control while reducing manual effort: DataClips~\cite{amini2016authoring} lets users compose, edit, and assemble clips from a reusable data-driven library to form complete data-video sequences; VisCommentator~\cite{chen2022viscommentator} supports rapid prototyping of augmented sports videos by combining machine-learning-based data extraction with visualization recommendations; WonderFlow~\cite{wang2025wonderflow} and Data Playwright~\cite{shen2025dataplaywright} provide narration-centric workflows for linking narration with visual elements and authoring instructions; Shen et al.~\cite{shen2025authoring} further support authoring data-driven chart animations through direct manipulation; Kineticharts~\cite{lan2022kineticharts} presents an affective animation design scheme for enhancing the expressiveness of charts in data stories; and Gemini$^2$~\cite{kim2021gemini2} supports keyframe-oriented chart-animation authoring by generating transitions between statistical graphics; see Section~\ref{subsec:declarative_specs}.}

\rev{\emph{Automated-generation systems} automate larger parts of the production workflow: InfoMotion~\cite{wang2021infomotion} automatically generates animated presentations from static infographics; AutoClips~\cite{shi2021autoclips} generates data videos from a tabular dataset and a pre-specified sequence of data facts by selecting, arranging, and configuring animated clips; Data Player~\cite{shen2024dataplayer} takes an existing visualization and text input, establishes semantic links between narration and visual elements with LLMs, and plans animation sequences through constraint solving; Narrative Player~\cite{shao2025narrativeplayer} takes a pre-written narrative paragraph paired with a corresponding data table and generates a coherent visualization sequence with transition animations and audio narration; and Shen et al.~\cite{shen2024datadirector} explore multi-agent workflows for automatic data-video creation.}

\rev{These systems differ in their input assumptions and design focus: some begin from existing visual artifacts, narration scripts, or pre-extracted data facts; others emphasize agentic generation workflows rather than an explicit shared declarative representation for cross-scene coordination. Recent empirical work~\cite{shen2025empirical} further examines how empirical findings have informed data-video creation tools, motivating the need to ground tool design in authors' workflows and creation needs. \system complements these approaches by starting from raw tabular data, jointly selecting and ordering scenes through multi-agent orchestration, and binding charts, narration, and animations in a shared declarative representation that makes cross-scene coherence and localized interactive editing explicit.}

\subsection{Automated Data Storytelling}
\rev{Automated data storytelling seeks to turn a data table into a narrative woven from data facts. Most existing work follows a data-fact-driven route: it extracts statistical facts from the table and then uses rules or search to select, order, and compose them into a story. DataShot~\cite{wang2020datashot} aggregates facts into fact sheets, Calliope~\cite{shi2021calliope} uses a logic-oriented search to find a coherent fact sequence, CoInsight~\cite{li2024coinsight} organizes connected insights in hierarchical tables, and Erato~\cite{sun2023erato} interpolates between user-specified facts to support collaborative editing. With the rise of large language models, recent work instead uses LLMs to generate the narrative and its visualizations directly; He et al.~\cite{he2025leveraging} survey how foundation models are applied across the stages of narrative visualization. For instance, DataNarrative~\cite{islam2024datanarrative} pairs a generator with an evaluator to produce stories that interleave text and visualizations, and InReAcTable~\cite{aodeng2025inreactable} turns this construction into an interactive process. Both lines of work, however, produce static output (fact sheets, documents, or chart sequences) that conveys insight through still graphics rather than motion and voice. \system targets a different medium: starting from raw tabular data, it decides which facts to tell and how to arrange them, and further produces multi-scene data videos that unify animation, narration, and temporal synchronization through a shared declarative representation.}

\subsection{Declarative Visualization and Animation Specifications}
\label{subsec:declarative_specs}
Declarative specifications have achieved widespread success in the visualization field by decoupling logical description from rendering implementation~\cite{chen2025chartmark,shen2022galvis,luo2024intelligent}.
D3~\cite{michael2011d3} pioneered the data-driven documents paradigm, while Vega-Lite~\cite{arvind2017vega} provides a concise grammar for interactive graphics.
In the animation domain, Canis~\cite{ge2020canis} designed a high-level language for chart animations, Gemini~\cite{kim2020gemini} provides a recommender system for animated transitions, \rev{and its successor Gemini$^2$~\cite{kim2021gemini2} further automates transition generation, Animated Vega-Lite~\cite{zong2022animated} unifies animation with the grammar of interactive graphics, and CAST~\cite{ge2021cast} and Data Animator~\cite{thompson2021data} support animation authoring through keyframes. These specifications excel at describing individual charts and chart animations, including transitions between chart states, but do not cover the multi-modal, multi-scene requirements of data videos that integrate charts, narration, and synchronized animation. \grammar extends them with data-driven semantic references for robustness and provenance, narration-indexed triggers for declarative audio-visual synchronization, and scene-level organization for multi-scene narratives.}

\section{Design Requirements}

\rev{Automatic data video generation requires coordinating data processing, visualization design, narration authoring, and animation configuration into coherent narrative sequences. Prior work on data video authoring workflows~\cite{shen2024dataplayer, amini2015understanding} and tool design~\cite{shen2025reflecting} reveals several recurring difficulties: cross-modal references that break upon content changes, narration--animation synchronization that depends on manual time alignment, the absence of scene-level structural organization, and the lack of systematic global narrative orchestration.}

\rev{In this work, we focus on multi-scene data videos generated from tabular data, in which insights are communicated through data-bound charts, narration, and synchronized animation. To clarify the practical scope of the system, we make explicit several design aspects: the chart forms available in the current implementation, the narrative structures that guide scene sequencing and script generation, and the shared visual settings and scene-specific layouts used across different scene types. Building on these observations and an analysis of the capability boundaries of existing tools, we derive four design requirements organized from lower-level representation to higher-level interaction.}

\textbf{DR1: Scene-driven modular organization.}
Prior research shows that effective data videos are typically composed of multiple semantically independent scenes, each conveying a distinct analytical insight~\cite{amini2015understanding,xie2026datamagic}. Workflow studies further indicate that authors naturally organize and edit content at the scene level during production~\cite{shen2024dataplayer}. However, pixel-level generation methods treat a video as a continuous stream of frames, with no explicit scene boundaries: intermediate results are opaque, and local edits require regenerating the entire video. The system should therefore treat scenes as the primary semantic unit of content organization, enabling each scene to be independently generated, rendered, and edited, thereby providing a foundation for modular content production and flexible narrative composition.

\textbf{DR2: Data-bound declarative audio-visual synchronization.}
Producing data videos requires establishing semantic references among narration, charts, and animated visual elements to maintain cross-modal consistency~\cite{Cheng2022Investigating,wang2025vizta}. Existing tools rely on manually specified DOM identifiers and absolute timestamps to implement such bindings~\cite{shen2024dataplayer, shen2025dataplaywright, ge2020canis}; these hard-coded references are fragile and break when data is updated, chart types are changed, or narration text is revised, leading to substantial rework. The system therefore needs a unified intermediate representation that (1)~references visual elements by data attribute values rather than rendering identifiers, ensuring references remain valid across data and chart changes while guaranteeing that every visual element is traceable to the underlying data; and (2)~replaces absolute timestamps with a declarative trigger mechanism, deferring temporal alignment to the render stage rather than requiring manual computation during authoring.

\textbf{DR3: Global narrative orchestration.}
Existing AI-assisted data video tools primarily support the generation of individual components (e.g., charts or scripts) while providing limited support for organizing multiple scenes into globally coherent narratives~\cite{shen2025reflecting}. High-quality data videos require accurate per-scene content and coherent narrative logic across scenes, such as a progression from macro to micro or from phenomenon to cause~\cite{yang2022design, amini2015understanding}. Greedy scene-by-scene generation easily produces information redundancy or narrative discontinuities. The system should therefore support global optimization over a candidate scene pool: selecting a scene subset based on insight value and query coverage, planning the playback order, and generating context-aware narration to ensure smooth and coherent scene transitions.

\begin{figure}[!t]
    \centering 
    \vspace{-0.5em}
    \includegraphics[width=0.9\linewidth]{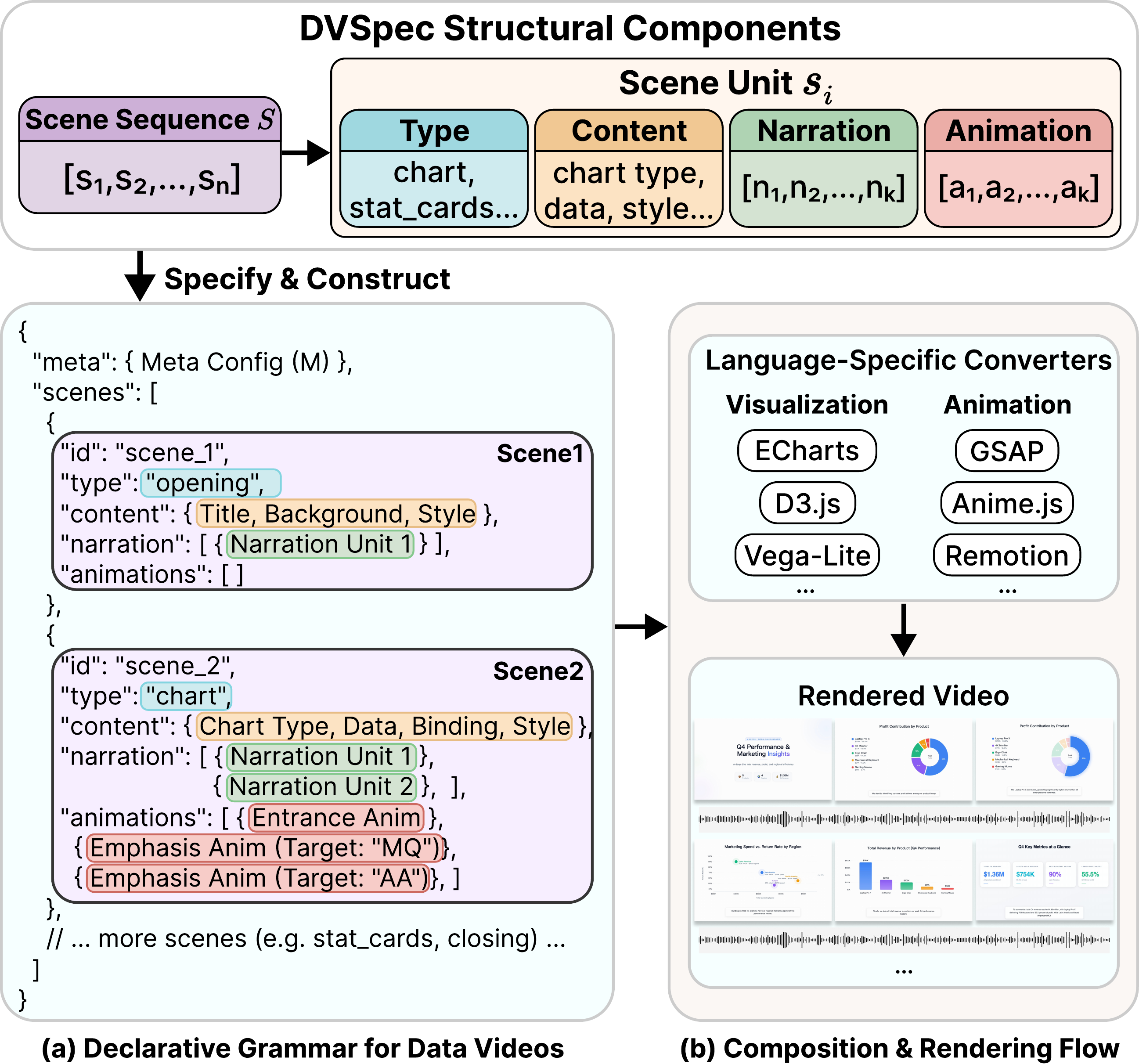}
    \vspace{-0.5em}
    \caption{\grammar structure and rendering flow.}
    \label{fig:dvspec_overview}
    \vspace{-2em}
\end{figure}

\textbf{DR4: Human-in-the-loop refinement via multi-modal interaction.}
Fully automated generation cannot anticipate all user preferences regarding visual style, narration tone, and analytical focus~\cite{tang2026vividoc,tang2026sketch,tang2026demonstrating}. Prior work on visualization and data-video authoring highlights that iterative refinement is central to real-world creation workflows~\cite{shen2026debugging,shen2024dataplayer,shen2025reflecting}. Moreover, different users favor different interaction modalities: some prefer direct manipulation of visual elements~\cite{luo2020visclean,luo2020interactive}, others prefer editing structured scripts~\cite{su2026vcgbench}, and still others prefer issuing natural language commands~\cite{li2024dawn,wu2024chartinsights}. The system should support complementary editing modalities over a shared representation, enabling users to switch among them without losing context and bridge full automation with fine-grained human control throughout the authoring process.

\section{\grammar: A Declarative Specification for Data Videos}\label{sec:grammar}

Existing declarative visualization specifications (e.g., Vega-Lite~\cite{arvind2017vega}, Canis~\cite{ge2020canis}) perform well for static charts or single-chart animations, but have not been extended to the unified description of cross-modal content and temporal coordination required for multi-scene data videos. To fill this gap, we propose \grammar (Data Video Specification), a declarative specification for data videos. Drawing on declarative design principles from the visualization grammar field~\cite{arvind2017vega, michael2011d3, shen2024dataplayer, shi2025virtualoverlays}, \grammar decouples logical description from rendering implementation to provide a unified intermediate representation.

\grammar takes scenes as its core organizational unit, decomposing a video into a self-contained scene sequence; within each scene, data-driven semantic references bind visual and animation elements to underlying data fields; animation trigger timing is declared through narration indices and resolved automatically at render time. As shown in Figure~\ref{fig:dvspec_overview}, \grammar encodes the video as a JSON object compiled by a language-agnostic renderer.

\subsection{Task Definition}\label{subsec:task_def}
We define data video generation as a mapping function from structured data to audiovisual narrative: $V = F(D, Q)$, where $D$ represents the dataset, $Q$ represents the user query, and $V$ is the generated data video. This task requires the system to understand the analytical intent in $Q$, extract data insights from $D$, and transform them into visual charts, narration scripts, and synchronized animations, ultimately composing coherent audiovisual content.

To address the temporal complexity of video generation, we represent the temporal content of $V$ as an ordered sequence of scenes: $S = [s_1, s_2, \ldots, s_n]$. Each scene $s_i$ is a self-contained semantic unit, formalized as a 5-tuple:
{
$$s_i := (id, type, content, narration, animation)$$
}
where \texttt{id} is a unique identifier, \texttt{type} defines the scene type (e.g., \texttt{chart}, \texttt{opening}, \texttt{stat\_cards}, \texttt{closing}), \texttt{content} encapsulates the visualization configuration, \texttt{narration} is a sequence of narration segments, and \texttt{animation} is a list of animation effects. This definition forms the basis for the three mechanisms described below.

\subsection{Scene-Driven Organization}\label{subsec:scene_org}

\grammar formalizes a data video as a structure containing metadata and a scene sequence: $V := (M, S)$. The metadata $M$ defines the global properties of the video (e.g., title, resolution), while $S$ is the ordered scene sequence defined in Section~\ref{subsec:task_def}, with each scene following the structure specified above.
\rev{At the video level, metadata provides global presentation settings shared across scenes; at the scene level, style and layout parameters in \texttt{content} express local visual organization matched to each scene's role. Together, these fields support a coherent visual identity across the video while allowing different scene roles, such as chart, opening, stat-card, and closing scenes, to use role-specific forms.}

Taking the ``Daily Revenue Trend'' scene in Figure~\ref{fig:grammar_example} as an example, its \texttt{content} field (\ding{172}) encapsulates the complete chart configuration, including the chart type (\texttt{line\_chart}), data binding rules (date on x-axis, revenue on y-axis), a data slice with 30 data points (\ding{173}), and visual style and layout parameters.

The \texttt{narration} field of a scene is an ordered list of narration segments $[n_0, n_1, \ldots, n_{k-1}]$ (\ding{174} in the figure). After text-to-speech (TTS) processing~\cite{azuretts}, each narration segment $n_j$ is assigned text content and a precise time range, formalized as:
{
$$n_j := (text, time\_start, time\_end)$$
}
The \texttt{animation} field of a scene is a list of animation effects $[a_0, a_1, \ldots, a_{m-1}]$. Each animation effect $a_\ell$ is formalized as a 4-tuple:
{
$$a_\ell := (type, target\_data, trigger, style)$$
}
where \texttt{type} is the animation type (e.g., \texttt{emphasis}), \texttt{target\_data} is a semantic reference, \texttt{trigger} is the trigger timing index, and \texttt{style} contains animation style parameters.

This scene-driven organization provides a modular and extensible foundation. By combining different scene types, the system can construct diverse narrative structures, while the self-contained nature of scenes facilitates parallel processing and independent rendering.

\subsection{Data-Driven Semantic Referencing}

\begin{figure}[!t]
    \centering 
    \includegraphics[width=0.9\linewidth]{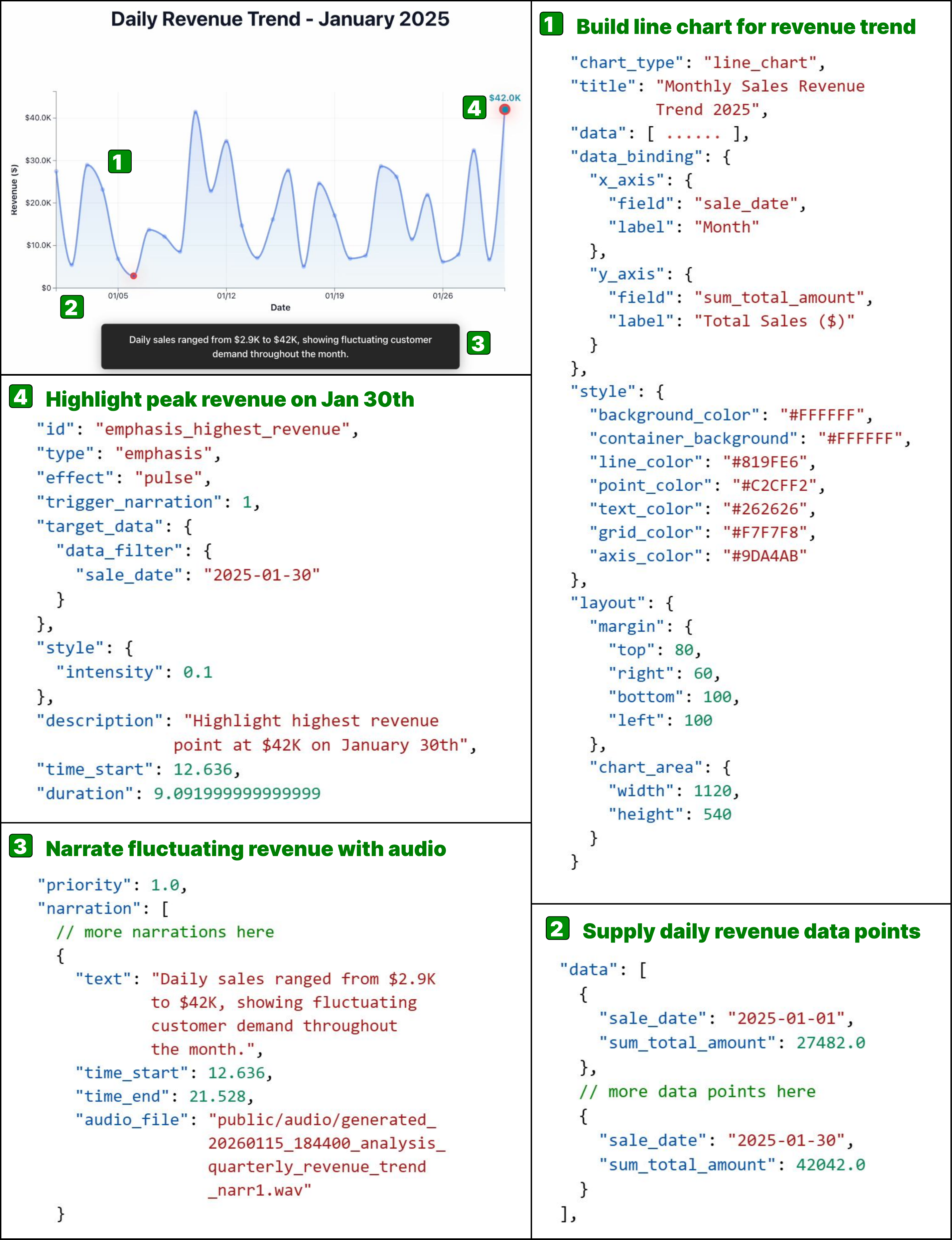}
    \vspace{-0.5em}
    \caption{A \grammar specification example for one scene, where the numbered markers (1--4) link each declarative field (chart, data, narration, and emphasis animation) to its effect in the rendered scene.}
    \label{fig:grammar_example}
    \vspace{-1.5em}
\end{figure}

\begin{figure*}[!t]
    \centering 
    \vspace{-1em}
    \includegraphics[width=0.85\linewidth]{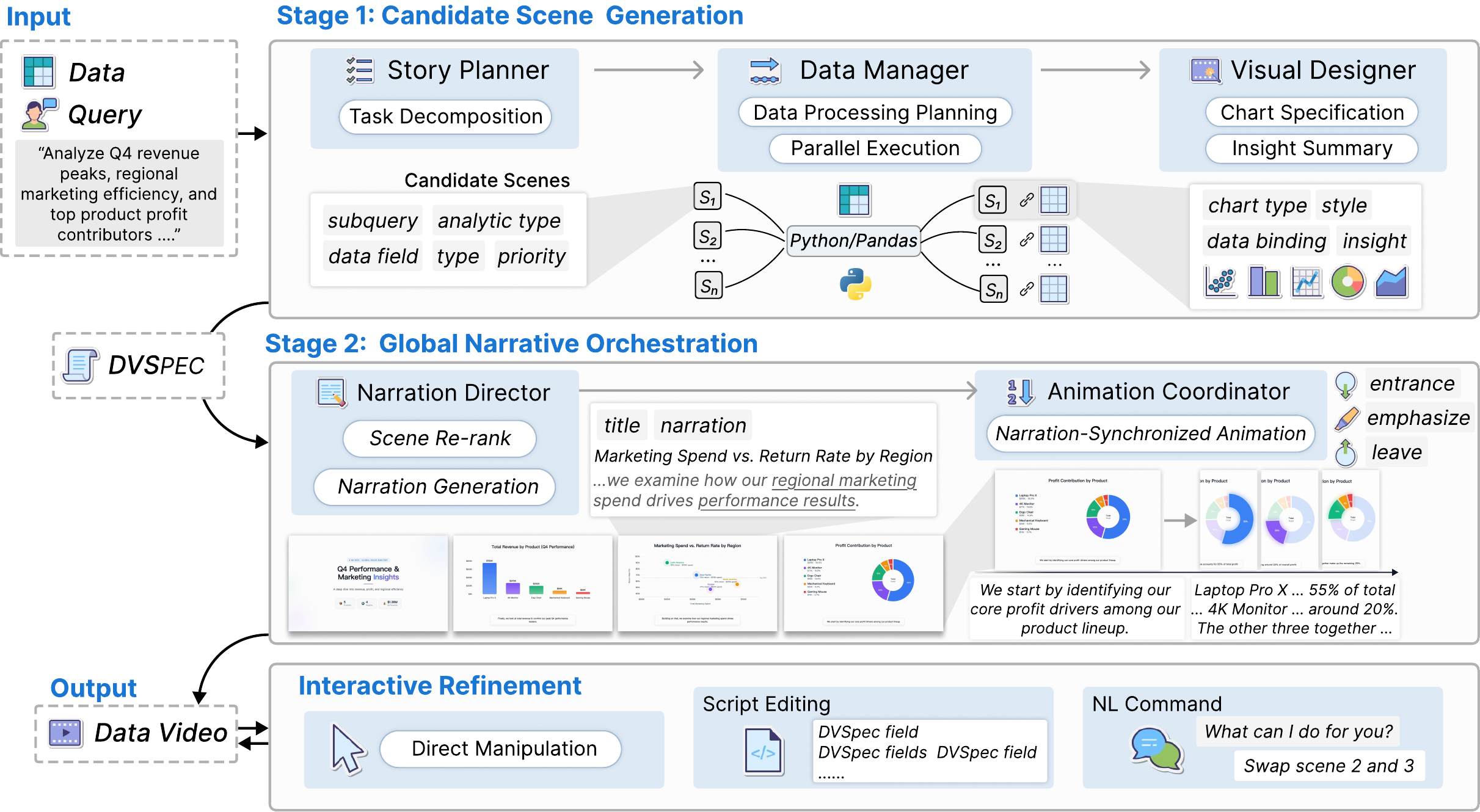}
    \vspace{-0.5em}
    \caption{System architecture of \system.}
    \label{fig:framework}
    \vspace{-1em}
\end{figure*}

In data videos, animation effects must precisely target specific visual elements, such as highlighting a data point on a particular date or emphasizing a specific bar. Traditional approaches use hard-coded rendering identifiers (e.g., DOM IDs) for this purpose. While workable in static settings, this is fragile in automated generation: changes in data ordering, chart type (e.g., from line to bar), or datasets may cause identifiers to point to wrong elements or become invalid.

\grammar adopts a semantic referencing mechanism based on data attribute values. Each animation's \texttt{target\_data} field specifies a set of key-value pairs (e.g., \texttt{\{"sale\_date":\,"2025-01-30"\}}) rather than rendering-layer identifiers. At render time, the system retrieves all records satisfying these conditions from the scene's dataset and binds the animation to the corresponding chart nodes.

As shown in Figure~\ref{fig:grammar_example} (\ding{175}), the animation references the January 30 data point via \texttt{\{"sale\_date":\,"2025-01-30"\}}. Regardless of data ordering or chart type changes, this reference always resolves to the correct visual element as long as the data structure is preserved. This mechanism provides two benefits: \textit{referencing robustness}, as the logical description is fully decoupled from rendering implementation without requiring the generation stage to anticipate rendering details; and \textit{data provenance}, as every visual element can be precisely traced back to its underlying data record, providing a structured foundation for the provenance-based interaction described in Section~\ref{sec:system_framework}.

\subsection{Narration-Indexed Declarative Triggering}

Audio-visual synchronization is the defining characteristic that distinguishes data videos from static visualizations. Traditional approaches require creators to manually specify an absolute timestamp for each animation. When narration text is revised or speaking rate changes, every timestamp must be recalculated, which is prohibitively costly in automated generation and iterative editing.

\grammar replaces absolute timestamps with declarative trigger relations. Given a scene's narration sequence $Narration = [n_0, n_1, \ldots, n_{k-1}]$, each animation's \texttt{trigger} field specifies the index of the narration segment that fires it ($trigger \in [0, k-1] \cup \{\text{null}\}$). If \texttt{trigger} is \texttt{null}, the animation executes at scene start; otherwise, it fires when the specified segment plays. At render time, the system automatically computes the precise trigger moment from the actual audio duration produced by TTS~\cite{azuretts}.

As shown in Figure~\ref{fig:grammar_example} (\ding{175}), \texttt{trigger:\,1} fires the animation when the second narration sentence (\ding{174}) plays. When the user revises the narration text, TTS regenerates the audio and trigger timing updates automatically, with no manual adjustment required.

This declarative mechanism separates the \textit{logical intent} of ``when to trigger'' (expressed via narration index) from the \textit{physical computation} of the exact trigger moment (resolved by the renderer). For automated generation, the multi-agent system only needs to specify the logical correspondence between animations and narration segments, without predicting final audio durations. For interactive editing, synchronization is automatically maintained after narration changes, eliminating the burden of manual timeline adjustment.

Together, the three mechanisms form the complete design of \grammar, providing a unified intermediate representation for both automated generation and interactive editing.

\section{System Framework}\label{sec:system_framework}

The generation of data videos involves complex dependencies among data processing, visual design, and narrative logic. Single-stage approaches struggle to ensure both per-scene accuracy and overall narrative coherence: sequential scene-by-scene generation locks in early decisions based on local information, leading to narrative repetition or coverage gaps; attempting to generate the full video at once faces combinatorial explosion in the design space.

Our core insight is that video generation can be decoupled into two relatively independent subproblems: \textit{scene content generation}, which focuses on accurately expressing individual analytical tasks, and \textit{global sequence orchestration}, which focuses on logical flow and temporal alignment across scenes. The orchestration targets video quality across five dimensions (Intent, Insight, Narrative, Animation, Aesthetic) subject to default constraints on total duration (60--120 seconds) and initial scene count ($k \leq 7$)~\cite{amini2015understanding}. Based on this, \system adopts a ``Generate-then-Orchestrate'' strategy implemented through two collaborative phases, as shown in Figure~\ref{fig:framework} and Algorithm~\ref{alg:datamagic}. In the candidate scene generation phase, the Story Planner, Data Manager, and Visual Designer collaborate to produce diverse candidate scenes in parallel. In the global narrative orchestration phase, the Narration Director and Animation Coordinator handle scene selection, ordering, and audio-visual synchronization. All agents collaborate through \grammar as a unified interface; detailed agent prompts are provided in \suppappendixref{C}{appendix:agent_prompts}.

\subsection{Candidate Scene Generation}

The goal of this stage is to generate a diverse candidate pool covering multiple analysis dimensions.

\textbf{Task Decomposition.} The Story Planner hierarchically decomposes the user query $Q$ (Algorithm~\ref{alg:datamagic}, Line~2), identifying key analytical dimensions from the dataset's field characteristics (e.g., temporal trends, categorical comparisons, geographic distributions) and splitting the high-level query into independent subtasks $\{q_i\}$. For example, the query ``analyze sales performance'' might be decomposed into ``revenue trend analysis,'' ``regional comparison,'' and ``product profit contribution.'' The decomposition follows an orthogonality principle: each subtask corresponds to an independent analytical perspective, maximizing the diversity of candidate scenes.

\begin{algorithm}[t!]
  \caption{\system Generation Process}
  \label{alg:datamagic}
  \begin{algorithmic}[1]
    \REQUIRE Query $Q$, Dataset $D$
    \ENSURE Video $V$
    \STATE // Candidate Scene Generation
    \STATE $tasks \gets \text{Planner.Decompose}(Q, D)$
    \STATE $pool \gets []$
    \FOR{each $q \in tasks$}
      \STATE $s \gets \text{InitScene}(q)$
      \STATE $s.data \gets \text{DataManager.Extract}(q, D)$
      \STATE $s.visual \gets \text{Designer.Design}(s.data)$
      \STATE $pool.\text{add}(s)$
    \ENDFOR
    \STATE // Global Narrative Orchestration
    \STATE $(S^*, \pi^*) \gets \text{Director.Select}(pool, Q)$
    \STATE $V \gets \text{InitVideo}()$
    \FOR{$j = 1$ {\bfseries to} $|S^*|$}
      \STATE $s \gets S^*[\pi^*[j]]$
      \STATE $ctx \gets \text{BuildContext}(S^*, \pi^*, j)$
      \STATE $s.narration \gets \text{Director.Narrate}(s, ctx)$
      \STATE $s.animation \gets \text{Animator.Sync}(s.vis, s.narr)$
      \STATE $V.\text{scenes}.\text{append}(s)$
    \ENDFOR
    \STATE {\bfseries return} $V$
  \end{algorithmic}
\end{algorithm}

\textbf{Data Extraction.} For each subtask $q_i$, the Data Manager plans a data processing workflow and automatically generates Python code to extract, filter, and aggregate relevant data from the source table, producing a micro-dataset $D_i$ targeted at that analytical objective (Line~6)~\cite{hong2025datainterpreter,li2026dataspace,li2026deepeye_workflow}. Separating data extraction from visualization design allows the data processing logic to be independently verified~\cite{datamosaic,datamosaic_demo}, and the same data slice can support exploration of multiple visualization schemes.

\textbf{Visualization Design.} The Visual Designer designs the visualization scheme based on $D_i$ (Line~7), selecting \rev{a chart type matched to the analytical task and data characteristics, configuring data bindings and style parameters, and extracting key statistical insights (e.g., extrema, trend directions, notable differences)~\cite{luo2022ncnet,luo2025nvbench2,luo2021synthesizing,qin2020making}. The current implementation centers on common data-bound statistical chart families, such as bar, line, area, scatter, pie, and heatmap charts; the available range is jointly shaped by the agent prompts, underlying model capabilities, and implemented rendering components, and can be broadened by extending the corresponding rendering components and generation rules.} \rev{These chart specifications and scene-level visual configurations} are written into \grammar's \texttt{type} and \texttt{content} fields. Notably, the \texttt{narration} and \texttt{animation} fields are intentionally left unfilled at this stage, keeping visualization generation decoupled from narrative construction; narration and animation require global context and are handled in the orchestration phase.

All candidate scenes are collected into the candidate pool (Line~8) for global optimization.

\subsection{Global Narrative Orchestration}

The candidate generation stage produces a pool of scenes that are individually coherent but narratively independent. This stage organizes the discrete candidate materials into a coherent audio-visual narrative through three steps: scene selection and ordering, context-aware narration generation, and audio-visual synchronization binding.

\textbf{Scene Selection and Ordering.} The Narration Director selects a subset $S^*$ from the candidate pool and plans the playback order $\pi^*$ (Algorithm~\ref{alg:datamagic}, Line~11). Selection is guided by two criteria: insight value (whether the scene reveals a meaningful pattern in the data) and query coverage (whether the selected scenes collectively address all aspects of the user query). \rev{Ordering and script generation follow a narrative pattern matched to the shape of the data story. The pattern guides scene ordering, climax placement, pacing, and transition style: Freytag's Pyramid by default and otherwise an inverted-pyramid, comparison-driven, time-driven, or drill-down structure. These patterns are consistent with the narrative structures studied in data storytelling~\cite{yang2022design, amini2015understanding}.}

\textbf{Context-Aware Narration Generation.} \rev{With the scene sequence and narrative pattern determined, the Director generates the opening, scene narrations, optional stat cards, and closing for the video (Line~16).} Rather than generating each scene's narration in isolation, the system uses a sliding window mechanism: the \texttt{BuildContext()} function constructs context from the current scene's position in the sequence (Line~15), incorporating the content and insights of adjacent scenes into the prompt. This allows the Director to generate narration that naturally bridges scenes. For example, when the preceding scene analyzed overall revenue trends, the next scene's narration can open with ``Let us now examine regional performance in detail'' rather than reintroducing background context. This mechanism helps maintain narrative continuity across scene transitions.

\begin{figure*}[!t]
  \centering 
  \vspace{-1.5em}
  \includegraphics[width=0.9\linewidth]{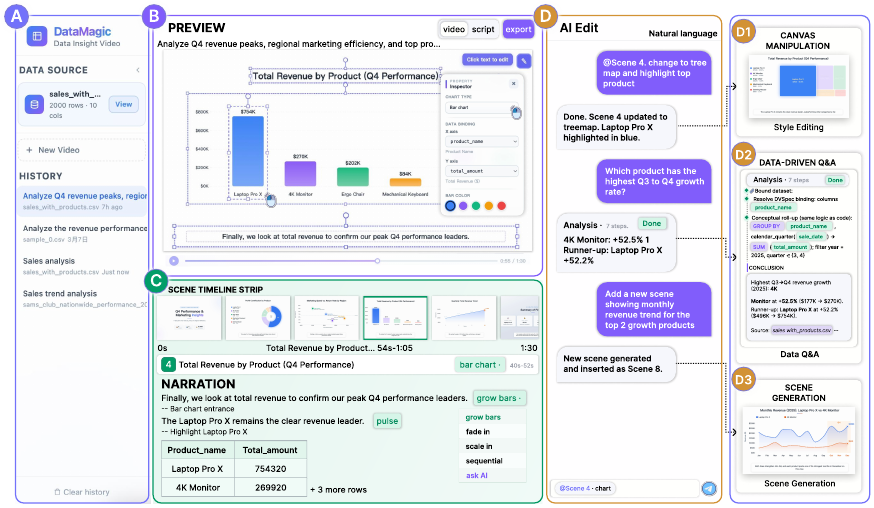}
  \vspace{-0.5em}
  \caption{\system web interface, showing (A--D) the main UI components and (D1--D3) three key interaction flows.}
  \label{fig:system_overview}
  \vspace{-1.5em}
\end{figure*}

\textbf{Audio-Visual Synchronization Binding.} The Animation Coordinator handles final audio-visual alignment (Line~17). It analyzes data entities mentioned in each narration segment (e.g., specific product names, dates, or values), uses \grammar's semantic referencing to bind these entities to corresponding visual elements, and specifies trigger timing via narration indices. For example, when the narration states ``Laptop Pro X contributed 55\% of revenue,'' the Coordinator generates an emphasis animation targeting \texttt{\{"product":\,"Laptop Pro X"\}} with its \texttt{trigger} set to the index of that narration segment. The visual highlight and voice narration are thus logically bound at the declarative level; the precise timestamp is automatically computed by the renderer from the TTS audio duration.

The complete \grammar configuration is compiled by the rendering engine into the final video, with narration text synthesized into speech via TTS~\cite{azuretts}. The system adopts a model-agnostic design supporting different LLM backends.

\subsection{Interactive System}

Built upon the generation framework above, we implement \system as a complete web-based interactive system (Figure~\ref{fig:system_overview}), developed with React~\cite{react} and Remotion~\cite{remotion}. \rev{The core design goal is to support human-in-the-loop refinement (DR4): users can inspect, understand, and selectively refine the generated video through a shared declarative \grammar state, without regenerating the entire video. \grammar organizes the video into self-contained scenes, allowing each interaction to be localized to a target scene and resolved within it without disturbing the rest of the video.}

The interface consists of four functional areas: the Data \& History Panel (A) for dataset upload and history browsing; the Preview \& Script Area (B) offering video preview, structured \grammar view, and export modes; the Scene Timeline \& Narration Editor (C) for direct narration and animation editing (with timing automatically re-aligned via \grammar's narration-indexed triggering); and the AI Edit Panel (D) providing a natural language interface. \rev{We first describe the design considerations and assumptions of the interactive system, and then the three interaction modes.}

\rev{\textbf{Design Considerations and Assumptions.} The interactive system follows a principle of \emph{scoped refinement}: each user request is interpreted as an update with an explicit scope over \grammar, rather than an uncontrolled rerun of the entire video-generation workflow. For within-scene changes, such as adjustments to chart type, data bindings, style parameters, narration, or animation triggers, the system updates the relevant \grammar fields in the target scene and re-renders only that scene, without rerunning the generation pipeline or automatically re-orchestrating the global narrative. Requests to reorder scenes or perform irreversible operations are first presented to the user for confirmation rather than applied silently. Users can express such requests through natural-language commands, structured script editing, or canvas actions; the system maps them to scoped modifications without requiring direct manipulation of the underlying grammar. By making the affected state and cross-modal bindings explicit and controllable (DR4), the system enables generated results to be refined in a predictable and user-steerable manner.}

\textbf{Canvas Manipulation (D1).} Users issue natural language instructions (e.g., \texttt{@Scene 4: change to treemap and highlight top product}) or click directly on the canvas to modify chart styles and data bindings. \rev{The system resolves such operations as a scoped modification of the target scene with incremental re-rendering, without full video regeneration; following the scoped-refinement principle above, the modification is confined within that scene and does not affect any other scene's configuration.}

\textbf{Data-Driven Q\&A (D2).} Users can pose data queries about the video content (e.g., ``Which product has the highest Q3 to Q4 growth rate?''). Rather than inferring answers from pixels, \system leverages \grammar's data provenance to enable structured queries. Specifically, the system first uses an LLM to parse the semantic intent of the question, then locates relevant data binding fields in the current scene's \grammar configuration, constructs structured query operations (e.g., filtering, aggregation, extremum retrieval), and executes them directly against the underlying tabular data. Results are returned with a full reasoning trace that records the data columns, filter conditions, and aggregation operations used, enabling verifiable and traceable responses.

\textbf{Scene Generation (D3).} Insights surfaced through Q\&A can be directly converted into new scenes (e.g., ``Add a scene showing the monthly revenue trend for the top 2 growth products''). The system automatically generates the corresponding \grammar configuration and inserts the new scene into the existing sequence, closing the loop from data exploration to narrative extension and transforming the data video from a one-way medium into an explorable interactive data interface.

\section{Experiments}\label{sec:experiments}

\begin{table*}[t!]
  \vspace{-1.5em}
  \caption{End-to-End Video Generation Performance Comparison.}
  \vspace{-0.5em}
  \label{tab:perf_comparison}
  \centering
    \begin{scriptsize}
      %\begin{sc}
      \setlength{\tabcolsep}{4pt}
        \begin{tabular}{lccccccc}
          \toprule
            &  
            & \multicolumn{6}{c}{\textbf{Evaluation Dimensions}} \\
            \cmidrule(lr){3-8}
          \textbf{Method} & Exec Rate (\%) & Intent & Insight & Narrative & Animation & Aesthetic & Avg. Score \\
          \midrule
          \multicolumn{7}{l}{\textbf{Direct Generation Methods}} \\
          DeepSeek-V3.2      & 48.62\% & 1.95 & 1.98 & 1.88 & 1.65 & 2.09 & 1.91 \\
          Gemini-2.5-Pro     & 66.06\% & 2.38 & 2.25 & 2.07 & 1.94 & 2.44 & 2.22 \\
          GPT-5              & 86.24\% & 2.36 & 2.22 & 2.05 & 1.84 & 2.17 & 2.13 \\
          Claude-Sonnet-4    & 84.40\% & 2.28 & 2.01 & 1.98 & 1.91 & 2.71 & 2.18 \\
          \midrule
          \multicolumn{7}{l}{\textbf{\system with Different Base Models}} \\
          \system (DeepSeek-V3.2)   & 96.33\% & 3.21 & 2.95 & 3.16 & 4.05 & 3.84 & 3.44 \\
          \system (Gemini-2.5-Pro) & 97.25\% & 3.28 & 3.17 & 3.56 & 3.89 & 3.44 & 3.47 \\
          \system (GPT-5)           & 95.41\% & 3.26 & 3.16 & 3.16 & 3.79 & 3.53 & 3.38 \\
          \system (Claude-Sonnet-4) & \textbf{98.17\%} & \textbf{3.79} & \textbf{3.37} & \textbf{3.84} & \textbf{4.39} & \textbf{4.05} & \textbf{3.89} \\
          \bottomrule
        \end{tabular}
      %\end{sc}
    \end{scriptsize}
  \vskip -0.1in
\end{table*}

\begin{table*}[!t]
  \caption{Ablation study of \system variants.}
  \vspace{-0.5em}
  \label{tab:autodv_ablation}
  \centering
    \begin{scriptsize}
      % \begin{sc}
      \setlength{\tabcolsep}{6pt}  % 列间距，可改为 8pt 更宽松
        \begin{tabular}{lcccccc}
          \toprule
          \textbf{Variant} & Intent & Insight & Narrative & Animation & Aesthetic & Avg. Score \\
          \midrule
          \textbf{\system (Full)}          & \textbf{3.79} & \textbf{3.37} & \textbf{3.84} & \textbf{4.39} & \textbf{4.05} & \textbf{3.89} \\
          w/o Story Planner      & 3.42 & 3.16 & 3.26 & 3.79 & 3.58 & 3.44 \\
          w/o Orchestration      & 3.32 & 3.21 & 3.42 & 3.95 & 3.79 & 3.54 \\
          \bottomrule
        \end{tabular}
      % \end{sc}
    \end{scriptsize}
  % \vskip -0.1in
  \vspace{-1.5em}
\end{table*}

\subsection{Experimental Setup}

\textbf{Evaluation Datasets.}
We collected evaluation data from two representative benchmarks: DAComp-DA~\cite{lei2025dacomp} contains real enterprise data from business scenarios such as sales and HR, while T2R-bench~\cite{zhang2025t2r} provides public statistics from 19 domains including environmental resources and public management. We selected single-table datasets and refined queries to suit video generation requirements. The final dataset consists of 60 datasets (36 from T2R-bench, 24 from DAComp-DA) with 109 samples. The data scale distribution includes small ($<$100 rows, 6 datasets), medium (100--1000 rows, 21 datasets), and large ($>$1000 rows, 33 datasets, up to 150,000 rows). The queries cover trend analysis, comparison, association, and distribution patterns. Detailed statistics are provided in \suppappendixref{B}{app:dataset}.

\textbf{Baselines.}
We establish two baseline categories: \textbf{(1) Direct Generation}, where GPT-5, Claude-Sonnet-4, Gemini-2.5-Pro, and DeepSeek-V3.2 each generate complete Remotion rendering code, including chart bindings, narration scripts, and animation configurations, through a single inference call, without any intermediate representation or multi-stage decomposition; \textbf{(2) \system Variants}, where the same LLMs are integrated into our framework to verify model-agnosticism. We exclude general video generators (e.g., Sora, Runway) as they cannot ensure the numerical accuracy and logical traceability required for data storytelling. Both direct generation baselines and \system variants use identical retry policies.

\textbf{Evaluation Metrics.}
We establish a two-tier evaluation framework: \textbf{(1) Coarse-grained metrics} include Execution Rate (Exec Rate) to assess system robustness; \textbf{(2) Fine-grained metrics} evaluate video quality across five dimensions on a 1--5 scale: \textit{Intent}, \textit{Insight}, \textit{Narrative}, \textit{Animation}, and \textit{Aesthetic}~\cite{bian2025you,xie2025visjudgebench,tang2026igenbench}. We employ Gemini-2.5-Pro as an automated evaluator. To validate its reliability, we performed stratified sampling of 60 videos and invited 3 experts with data video research backgrounds to independently score them using a unified rubric (1--5 scale), with averaged expert scores as ground truth. Automated scores exhibit strong positive correlation with expert scores overall (Pearson $r = 0.91$, $p < 0.001$; $MAE = 0.38$), with all dimension-level correlations exceeding 0.75. \rev{\evaluationappendixstatement}

\subsection{End-to-End Video Generation Performance}

As shown in Table~\ref{tab:perf_comparison}, \system significantly outperforms direct generation methods across all evaluation dimensions, with consistent improvements across different base models.

\textbf{Quality Improvement.} Direct generation methods achieve average scores between 1.91 and 2.22; \system improves these to 3.38--3.89. The two most significant gains are in Animation and Narrative. Animation scores increase from an average of 1.84 to 4.03 (+120\%): direct generation methods lack structured audio-visual alignment mechanisms, causing animation triggers to frequently desynchronize from narration content; \grammar's narration-indexed triggering resolves this at the declarative level. Narrative scores improve from 2.00 to 3.43 (+72\%): without global narrative planning, direct generation produces scenes with logical discontinuities or information repetition; the Generate-then-Orchestrate strategy substantially improves coherence through global orchestration. Intent and Aesthetic dimensions also improve by more than 50\%, while Insight improves by 49.5\%.

\textbf{Execution Stability.} Direct generation methods show high variance in execution rates (48.62\%--86.24\%), as errors at any stage cause complete generation failure. \system stabilizes execution rates above 95\% through its decoupled design. Notably, DeepSeek-V3.2 improves its execution rate from 48.62\% to 96.33\%, with its average score (3.44) surpassing the best-performing direct generation model, validating the framework's model-agnosticism.

\textbf{Efficiency Trade-off.} Direct generation methods average approximately 57 seconds, while \system's two-stage process totals approximately 176 seconds (60 seconds for configuration generation and 116 seconds for video rendering with parallelism set to 3), which can be further reduced by increasing parallelism. Although \system takes roughly three times longer, this results in significantly higher execution rates (95\%+ vs. 49\%--86\%) and quality (3.89 vs. 2.22), reflecting a deliberate trade-off of time for quality and robustness. Training a dedicated data-video model offers a potential path to reducing this additional overhead and reliance on external APIs~\cite{lin2026lead,xie2025visjudgebench,wu2026autowebworld}.

\subsection{Ablation Study}

To validate the effectiveness of key modules, we conducted ablation experiments using Claude-Sonnet-4 as the base model. We ablate the Story Planner and Orchestration module, the core designs that distinguish \system from direct generation, while excluding the Data Manager and Visual Designer, whose removal causes pipeline failure rather than measurable quality degradation.

As shown in Table~\ref{tab:autodv_ablation}, removing the Story Planner reduces the average score from 3.89 to 3.44 (11.6\% decrease), with Narrative and Animation showing the most significant drops (15.1\% and 13.7\%). This indicates that the Story Planner's role goes beyond query decomposition: without it, the system tends to generate candidate scenes with limited analytical diversity, compressing the optimization space available to the orchestration stage.

Removing the Orchestration module has a greater impact, reducing the average score to 3.54 (9.0\% decrease). Intent is most affected (12.4\% decrease) and Narrative also drops significantly (10.9\%), indicating that without global orchestration the system degrades to greedy scene-by-scene generation, making it difficult to ensure overall query coverage and cross-scene logical coherence.

Notably, even without Orchestration, the Animation dimension maintains a relatively high score (3.95, only 10.0\% decrease). This validates the robustness of \grammar's narration-indexed triggering: audio-visual synchronization is guaranteed at the declarative level by \grammar and does not depend on global orchestration. Together, the two ablations confirm the necessity of both Generate-then-Orchestrate stages.

\subsection{Error Pattern Analysis of Direct Generation}

To understand why direct generation methods perform poorly on the data video task, we systematically analyze their outputs and identify three categories of core error patterns.

\textbf{Visual Rendering Failures.}
This is the most severe error category, representing a fundamental breakdown where the video stream degenerates into a prolonged pure-color background and no longer conveys any data or narrative information. For instance, DeepSeek-V3.2's output turned completely white after 17 seconds. The root cause is execution failure in the code generation stage: rendering code produced by the model contains syntax errors or runtime exceptions that prevent visualization from being displayed.

\textbf{Visualization Design Defects.}
Even when content is successfully rendered, outputs frequently violate established visualization guidelines. The most prevalent issue is \textit{truncation and cropping}: models fail to adapt visual elements to canvas boundaries, causing titles or axis labels to be cut off, as observed in GPT-5's outputs. \textit{Element overlap} also occurs frequently, with text annotations or UI components occluding critical data points in violation of the non-occlusion principle. Additionally, models exhibit \textit{failed narrative animation}: rather than leveraging the temporal dimension of video to show trends or process evolution, they revert to static images, reducing data storytelling to decorative slideshows.

\begin{figure*}[!t]
    \centering
    \vspace{-1em}
    \includegraphics[width=1.0\linewidth]{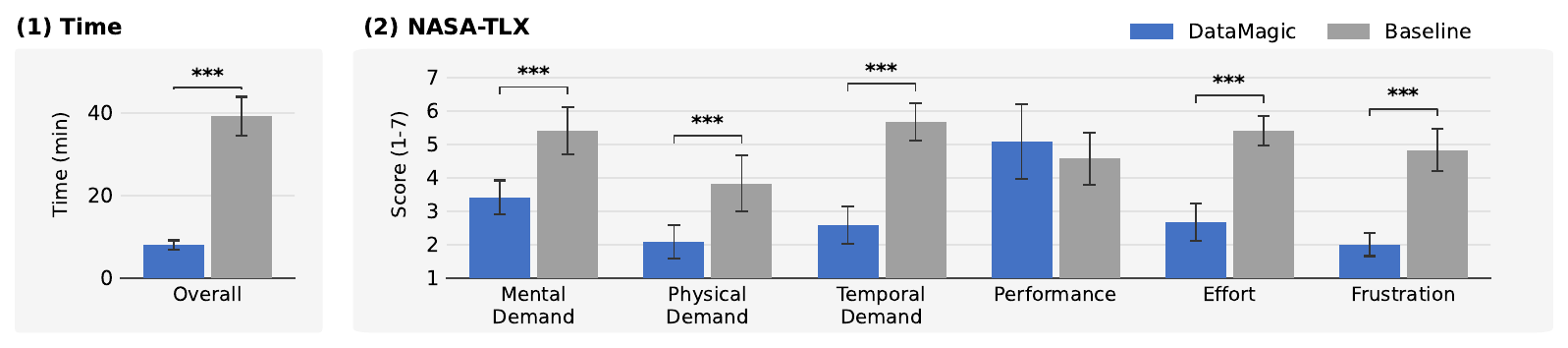}
    \vspace{-2em}
    \caption{\rev{User-study results ($N=12$). \system reduced task time from 39.2 to 8.0 min (79.7\%) and lowered ratings on five NASA-TLX workload dimensions, with no significant Performance difference. Error bars: 95\% CIs; Wilcoxon signed-rank tests: $^*p<.05$, $^{**}p<.01$, $^{***}p<.001$.}}
    \label{fig:userstudy}
    \vspace{-1.5em}
  \end{figure*}

\textbf{Audio-Visual Consistency Errors.}
The third category concerns semantic and temporal misalignment between visual and auditory channels. A recurring issue is \textit{temporal misalignment}: the narrator analyzes a trend while the screen remains black or lags on the previous scene, producing a ``blind narration'' effect observed in Gemini-2.5-Pro's outputs. The root cause is that direct generation methods lack explicit audio-visual binding mechanisms: narration and visual content are generated in a single inference pass, but the model cannot precisely control their temporal correspondence. \system eliminates this problem by design through \grammar's declarative triggering.

These three error patterns share a common underlying cause: end-to-end data video generation requires cross-modal coordination that exceeds the capability of current LLMs in a single inference pass.
\rev{\suppappendixref{A}{appendix:case_studies} provides qualitative case studies, covering a cross-method comparison of these error patterns (\suppappendixsubref{A.1}{subsec:appendix_comparison}), high-quality \system generations (\suppappendixsubref{A.2}{subsec:case_study}), and low-quality examples that analyze typical failure cases (\suppappendixsubref{A.3}{subsec:low_quality_cases}).}

\section{User Study}

To evaluate the practical utility of \system in real-world data video creation workflows, we conducted a within-subjects user study following the evaluation paradigm of recent HCI system research~\cite{wang2026tabletale,shen2025dataplaywright}.
\rev{The study had two evaluation goals: first, to assess whether \system improves video creation efficiency and reduces cognitive load relative to a conversational LLM workflow; and second, to understand how users perceive its usability and interactive editing capabilities.}

\subsection{Study Design}

\textbf{Conditions.}
Each participant completed one task under each of two conditions, with condition order counterbalanced:
(1)~\textit{\system}: participants used the web-based system proposed in this paper, triggering the multi-agent generation pipeline via a natural language query and iteratively editing the output using the three interaction modes described in Section~\ref{sec:system_framework};
(2)~\textit{Conversational LLM workflow (baseline)}: participants used the same underlying LLM (Claude Sonnet 4.5) to perform data analysis, narration scripting, and visualization code generation through multi-turn dialogue, then rendered the video using a pre-configured Remotion environment.
We chose this baseline over existing authoring tools (e.g., DataClips~\cite{amini2016authoring}, Data Player~\cite{shen2024dataplayer}), which require pre-prepared visualizations and thus cannot support our end-to-end, raw-data-to-video evaluation setting.
The baseline shares the same technology stack as \system (Claude Sonnet 4.5 and Remotion); the key difference is the absence of \grammar's structural constraints and the multi-agent orchestration layer, requiring participants to coordinate each step manually and manage audio-visual synchronization by hand.
\rev{Both conditions had a 40-minute target duration to simulate a time-pressured creation environment~\cite{wang2026tabletale}. Participants who exceeded this target duration were allowed to continue until completion, and their actual completion times were recorded.}

\textbf{Participants.}
We recruited 12 participants (5 male, 7 female; aged 21--29) through an open call at a local university (P1--P12). Participants were Master's or junior Ph.D. students from diverse fields (computer science, data science, finance, business analytics), all with data analysis experience. Based on self-reported experience with data video creation, we divided them into two groups: 6 novices with limited experience in video editing or data storytelling, and 6 experts with prior experience creating data visualizations or videos. All participants were fluent in English and provided written informed consent before participating in the study. The study was approved by the HKUST Human and Artefacts Research Ethics Committee (No. HKUST(GZ)-HSP-2026-0196). Upon completion, each participant received cash compensation equivalent to USD~10 for their time.

\textbf{Tasks.}
To cover representative data video scenarios and enhance ecological validity, we designed three types of narrative tasks:
(1)~\textit{Attribution}: explain the key drivers behind a metric change;
(2)~\textit{Evolution}: analyze temporal trends and inflection points in a time series;
(3)~\textit{Comparison}: evaluate differences across multiple entities on various dimensions.
Each participant was assigned different task types across the two rounds. Every task covered the full creation workflow, including data analysis, visualization design, narration writing, and animation configuration, and required at least one editing operation to assess interactive editing capability.

\textbf{Ordering and Procedure.}
We used a Balanced Latin Square design to ensure that task type, condition, and presentation order were evenly distributed across participants, with expertise level and condition order counterbalanced within each group. Each session lasted approximately 90--110 minutes and consisted of four phases: (1)~introduction and training ($\sim$10 min); (2)~first trial (40-min target duration) followed immediately by post-condition questionnaires; (3)~second trial under the other condition and task type; (4)~semi-structured interview ($\sim$10 min).

\textbf{Measures.}
We collected three types of data:
(1)~\textit{Objective}: task completion time per condition;
\rev{(2)~\textit{Subjective}: the Raw NASA-TLX to assess perceived cognitive load, and the System Usability Scale (SUS) item set to assess perceived usability, both administered on a 7-point Likert scale immediately after each trial to minimize recall bias; after reverse-coding negatively worded items, we report the mean per-item usability rating across the ten SUS items (1--7, higher is better), rather than the canonical 0--100 SUS composite;}
(3)~\textit{Task-specific feedback}: a custom 5-item, 7-point Likert questionnaire covering visual quality (Q1), animation effectiveness (Q2), narrative coherence (Q3), controllability (Q4), and overall satisfaction (Q5).
As questionnaire data did not satisfy normality assumptions, all subjective measures were analyzed using Wilcoxon signed-rank tests, with effect sizes reported as $r = z / \sqrt{N}$ ($N = 12$).

\subsection{Quantitative Results}

\rev{Overall, \system shows significant gains in task time, perceived usability, and most workload dimensions, providing converging evidence for both evaluation goals.}

\textbf{Task Completion Time.}
As shown in Figure~\ref{fig:userstudy}(1), \system significantly reduced task completion time (Wilcoxon $z = -3.06$, $p < .001$, $r = 0.88$).
Average completion time decreased from 39.2 minutes under the baseline to 8.0 minutes with \system, an improvement of approximately 79.7\%.
\rev{Notably, 7 participants (58.3\%) exceeded the 40-minute target duration under the baseline condition, while all 12 participants completed the \system condition within the target duration, indicating more predictable task completion.}

\textbf{System Usability.}
\rev{Using the SUS item set administered on a 7-point Likert scale, with negatively worded items reverse-coded, \system achieved a significantly higher mean per-item usability rating ($M = 5.97$, $SD = 0.35$) than the baseline ($M = 3.22$, $SD = 0.91$; $p < .001$). Ratings range from 1 to 7, with higher values indicating better perceived usability for the authoring workflow.}

\textbf{Cognitive Load.}
\rev{As shown in Figure~\ref{fig:userstudy}(2), \system significantly reduced perceived load on five of the six NASA-TLX dimensions (Mental Demand, Physical Demand, Temporal Demand, Effort, and Frustration; all $p < .001$), while Performance showed no significant difference ($p = .34$), indicating that the reduction in workload was not accompanied by a detectable difference in participants' self-rated task performance between conditions.}

\textbf{Task-Specific Feedback.}
\system outperformed the baseline on all five Likert dimensions.
The largest gaps were in narrative coherence (6.6 vs.\ 2.8) and animation effectiveness (6.3 vs.\ 3.0), directly corroborating the quantitative improvements in Animation and Narrative reported in Section~\ref{sec:experiments}.
Visual quality also showed a substantial gap (6.3 vs.\ 3.2), and overall satisfaction was significantly higher (6.4 vs.\ 3.2).
The controllability dimension showed the smallest difference (6.2 vs.\ 3.7), as the baseline's multi-turn dialogue itself provides considerable interaction flexibility.

\subsection{Qualitative Feedback}

Interview feedback revealed four themes consistent with the quantitative results.

\textbf{End-to-end automation eliminates workflow fragmentation.}
Under the baseline, each step required a separate dialogue round with manual handoff to the next. P3 noted: \textit{``It took me several rounds of chatting to get a usable chart, but then for narration and animation I had to start all over again.''} P7 similarly found that modifying chart data bindings required tediously re-synchronizing narration and animation by hand.

\textbf{Audio-visual synchronization becomes automatic.}
Multiple participants (P2, P5, P7, P9) identified timeline alignment as the most difficult aspect of the baseline. \grammar's narration-indexed triggering was seen as fundamentally resolving this. P9 noted: \textit{``The narration and animation just naturally fit together---I didn't have to align the timing myself.''}

\textbf{\rev{Interactive editing is effective and alleviates trust concerns about automation.}}
\rev{In the \system condition, each participant completed at least one intended edit; all 12 successfully applied their requested changes within the target scene without disturbing the rest of the video. This editability also reshaped attitudes toward automation: participants (P1, P4, P10) initially had reservations about automated outputs, but the three interaction modes effectively resolved their concerns. P11 commented: \textit{``Knowing I can still go in and adjust things actually made me more comfortable letting it generate automatically.''} Novices preferred natural language commands while experts favored direct script editing, confirming that the modes accommodate different user experience levels.}

\textbf{Suggestions for improvement.}
Participants suggested a scene template library (P6) and support for brand visual guidelines in export (P12), providing directions for future iterations.

\section{Discussion}
\label{sec:discussion}

\rev{\textbf{Expressiveness and practical scope.}
Although \grammar organizes charts, narration, animation, and scene structure within a unified specification, \system is designed to be extensible beyond its current set of chart and narrative forms. Its practical expressive range is jointly determined by the visual vocabulary implemented by the rendering engine and the analytical capabilities of the underlying models. Both layers can evolve: new visual forms can be incorporated by extending the rendering components, while stronger models can support more complex analyses. \system's practical expressiveness can therefore grow as these layers improve, while \grammar remains a stable core abstraction. This design helps preserve reliable chart rendering and traceability of visualized values to source data, while leaving room for future capability expansion.}

\rev{\textbf{Designing for user intervention.}
Fully automated generation cannot always anticipate users' analytical focus, visual style, or narration tone. \system therefore exposes \grammar as a shared, inspectable state rather than treating generation as a black box, enabling refinement through natural-language commands, structured script editing, and canvas actions. We adopt \emph{scoped refinement}: an edit updates only the relevant fields of its target scene and re-renders that scene, preserving cross-modal bindings without rerunning the full pipeline or silently altering user-authored decisions in other scenes. Because \grammar separates logical description from rendering implementation, these updates remain renderer-agnostic while keeping the specification and rendered output synchronized. This intervention, however, still occurs after generation. A promising next step is to present the analysis plan before full synthesis so users can confirm or adjust early decisions about analytical direction or audience framing before they propagate through the video.}

\rev{\textbf{Balancing animation reliability and creative diversity.}
\system uses structured animation patterns in the agent prompts, such as entrance effects and narration-triggered emphasis effects, to keep animations aligned with chart semantics and synchronized with the narration. This keeps the animation stable, legible, and in service of the data narrative. The same reliability, however, comes at the cost of diversity: the guidelines steer the system toward a conservative motion language and may suppress the more expressive animations an unconstrained model might attempt. Guideline-free prompting could yield more creative videos, but at present it would make legibility, pacing, and alignment with the narration harder to control consistently. A promising direction is to treat the guidelines as soft constraints that a model may depart from when it can justify the outcome, or to learn animation styles from curated exemplars.}

\rev{\textbf{Limitations and future work.}
The discussion so far has centered on \system's capabilities and design trade-offs; beyond these, the system has several concrete limitations, each pointing to a clear future direction. On the data side, \system targets single-table inputs, which already cover most data-storytelling scenarios; supporting multi-table joins and richer relational data would broaden its applicability. On the visual side, integrating image-generation models could supply non-data assets such as custom illustrations and further enrich the visuals. On evaluation, the data-video field still lacks a commonly accepted benchmark; building public benchmarks for reproducible cross-system comparison is an important step for the area as a whole. On rendering, \system currently binds \grammar to a single renderer; generalizing specification--rendering synchronization across different backends would let the renderer-agnostic specification drive diverse renderers and support rendering-specific edits.}

\section{Conclusion}

This paper presents \system, an end-to-end system that generates data videos from raw tabular data through declarative multi-agent orchestration. \grammar provides a shared state for agent collaboration and user editing while organizing charts, narration, and animation; the ``Generate-then-Orchestrate'' strategy decouples local scene generation from global narrative organization. Quantitative evaluations show that, compared with direct generation baselines, \system improves video quality and execution stability; the user study further shows that it increases authoring efficiency and reduces cognitive load. These results suggest that declarative representations can serve not only as output descriptions but also as an intermediate layer coordinating automated generation, data traceability, and fine-grained human control, thereby supporting structured and controllable authoring workflows.

\ifshowack
\acknowledgments{%
  This paper was supported by the National Science and Technology Major Project (2025ZD0619402); the NSF of China (62402409); Youth S\&T Talent Support Programme of Guangdong Provincial Association for Science and Technology (SKXRC2025461); the Young Talent Support Project of Guangzhou Association for Science and Technology (QT-2025-001); Guangzhou Basic and Applied Basic Research Foundation (2026A1515010269, 2025A04J3935, 2023A1515110545); and Guangzhou-HKUST(GZ) Joint Funding Program (2025A03J3714).%
}
\fi

\section*{Supplemental Materials}
\label{sec:supplemental_materials}

The supplemental materials include an appendix with qualitative cases, dataset statistics, agent prompts, evaluation criteria, and expert-validation procedures; a system demonstration video showing the end-to-end data-video authoring and interactive refinement workflow; and representative generated outputs that illustrate the quality and diversity of \system across different datasets and queries.

% \section*{Figure Credits and Copyrights}
% \label{sec:figure_credits}
% 
% All figures in this paper were created by the authors and are released under a \href{https://creativecommons.org/licenses/by/4.0/}{CC BY 4.0 license}.
% All figures remain under the authors' own copyright.

\bibliographystyle{abbrv-doi-hyperref}

\bibliography{reference}

\ifdefined\CameraReadyMain
\else
\clearpage

\appendix
\crefalias{section}{appendix}

\fi

\fi

\end{document}